\documentclass{article}

\usepackage[english]{babel}
\usepackage[utf8x]{inputenc}
\usepackage[T1]{fontenc}

\usepackage[a4paper,top=3cm,bottom=2cm,left=3cm,right=3cm,marginparwidth=1.75cm]{geometry}

\usepackage{authblk}

\usepackage{algorithm,algpseudocode}
\usepackage{caption}
\usepackage{bm}
\usepackage{amsmath,amssymb}
\usepackage{graphicx}
\usepackage{cleveref}
\usepackage{subfig}
\usepackage{physics}
\usepackage{tabularx}
\usepackage{threeparttable}
\usepackage{placeins}
\usepackage{jabbrv}

\Crefname{figure}{Fig.}{Figs.}

\usepackage{tikz}
\usetikzlibrary{shapes,arrows,matrix,calc,fit}

\newcommand{\OO}{\mathcal{O}}
\newcommand{\PP}{\mathcal{P}}

\newcommand{\FF}{\bm{\mathcal{F}}}
\newcommand{\SSj}{\bm{\mathcal{S}}_j}
\newcommand{\sOO}{{_{\OO}}}
\newcommand{\sPP}{{_{\PP}}}

\newcommand{\rr}{\bm{r}}
\newcommand{\sumj}{\sum_{j=1}^J}
\newcommand{\pgrad}{\boldsymbol{\partial}}
\newcommand{\maxnorm}[1]{\left\|#1\right\|_{\text{max}}}

\newcommand{\Rel}[1]{\mathfrak{R}[#1]}
\newcommand{\Img}[1]{\mathfrak{I}[#1]}
\newcommand{\Reals}{\mathbb{R}}
\newcommand{\Complexes}{\mathbb{C}}

\newcommand{\RO}{{\Rel{\OO}}}
\newcommand{\IO}{{\Img{\OO}}}

\newcommand{\delP}{\pgrad_{\sPP}}
\newcommand{\delO}{\pgrad_{\sOO}}
\newcommand{\delRO}{\pgrad_{_\RO}}

\newcommand{\ki}{\mathbf{k}_i}
\newcommand{\kf}{\mathbf{k}_f}

\tikzstyle{fwd} = [rectangle, draw, fill=blue!20, text centered, 
align=center,  minimum height=2em]
\tikzstyle{rev} = [rectangle, draw, fill=red!20, text centered,
align=center, minimum height=2em]
\tikzstyle{var} = [ellipse, draw, fill=black!10, text centered,
align=center, minimum height=2em]
\tikzstyle{line} = [draw, -latex', line width=0.35mm]

\tikzstyle{fwdc} = [ellipse, draw, fill=blue!20, text centered, 
align=center,  minimum height=2em]
\tikzstyle{revc} = [ellipse, draw, fill=red!20, text centered,
align=center, minimum height=2em]

\newcommand{\tmath}[1]{\begin{aligned}#1\end{aligned}}

\algnewcommand\algorithmicinit{\textbf{Initialize:}}
\algnewcommand\Initialize{\item[\algorithmicinit]}

\newcounter{parentalgorithm}

\makeatletter
\newenvironment{subalgorithms}{%
	\refstepcounter{algorithm}%
	\protected@edef\theparentalgorithm{\thealgorithm}%
	\setcounter{parentalgorithm}{\value{algorithm}}%
	\setcounter{algorithm}{0}%
	\def\thealgorithm{\theparentalgorithm\alph{algorithm}}%
	\ignorespaces
}{%
	\setcounter{algorithm}{\value{parentalgorithm}}%
	\ignorespacesafterend
}
\makeatother

\begin{document}
	
\title{Using automatic differentiation as a general framework for ptychographic reconstruction}
\author[1]{Saugat Kandel}
\author[2]{S. Maddali} 
\author[3]{Marc Allain}
\author[2]{Stephan O Hruszkewycz}
\author[4,5,6]{Chris Jacobsen}
\author[7,*]{Youssef S G Nashed}

\affil[1]{Applied Physics, Northwestern University, Evanston, Illinois 60208, USA}
\affil[2]{Materials Science Division, Argonne National Laboratory, Lemont, IL 60439, USA}
\affil[3]{Aix Marseille Univ, CNRS, Centrale Marseille, Institut Fresnel, Marseille, France}
\affil[4]{Advanced Photon Source, Argonne National Laboratory, Lemont, Illinois 60439, USA}
\affil[5]{Department of Physics \& Astronomy, Northwestern University, Evanston, Illinois 60208, USA}
\affil[6]{Chemistry of Life Processes Institute, Northwestern University, Evanston, Illinois 60208, USA}
\affil[7]{Mathematics and Computer Science Division, Argonne National Laboratory, Lemont, Illinois 60439, USA}

\affil[*]{Corresponding author: ynashed@anl.gov}
\date{}

\maketitle

\begin{abstract}
Coherent diffraction imaging methods enable imaging beyond lens-imposed resolution limits.  In these methods, the object can be recovered by minimizing an error metric that quantifies the difference between diffraction patterns as observed, and those calculated from a present guess of the object. Efficient minimization methods require analytical calculation of the derivatives of the error metric, which is not always straightforward. This limits our ability to explore variations of basic imaging approaches.  In this paper, we propose to substitute analytical derivative expressions with the automatic differentiation method, whereby we can achieve object reconstruction by specifying only the physics-based experimental forward model. We demonstrate the generality of the proposed method through straightforward object reconstruction for a variety of complex ptychographic experimental models.
\end{abstract}

\section{Introduction}
Ptychography is a coherent diffraction imaging (CDI) technique that acquires a series of
intensity diffraction patterns through spatial shifts of the illumination
(probe) across the sample (object) in a set of overlapping beam positions. 
Given a large number of overlapping beam positions, 
the ptychography experiment yields sufficient redundant information with which we can reconstruct the object structure to sub-beam-size spatial resolution, 
and even determine additional experimental parameters such as the structure of the probe itself. 
First proposed by Hoppe in 1969 \cite{hoppe_aca3_1969}, 
the ptychographic technique was realized experimentally 
and rapidly developed algorithmically in the 2000s
\cite{rodenburg_apl_2004,faulkner_prl_2004,guizar_oe_2008,maiden_ultramic_2009,thibault_ultramic_2009}.
By removing the typical CDI limitation that the probe size has to be larger than the sample, 
ptychography has enabled high resolution imaging of extended objects, 
making it a powerful imaging technique.
As such, the ptychographic technique has found application
not only as a 2D far-field diffraction imaging method but also as 2D near-field 
diffraction imaging method \cite{stockmar_scirep_2013}, a 3D Bragg imaging method \cite{godard_natcomm_2011}, 
a 3D multislice imaging method \cite{maiden_josaa_2012}, 
and a part of the 3D tomographic imaging method
\cite{dierolf_nature_2010} including for objects beyond the depth of
focus limit \cite{gilles_optica_2018}.

Imaging with ptychography involves solving the challenging phase retrieval problem, 
where one attempts to reconstruct an object from only the magnitude of its Fourier transform. 
In general, the phase retrieval problem is ill-posed \cite{bates_optik_1982};
solving it requires the use of oversampling and support constraints. 
These are typically used in an iterative projection framework that updates the object guess
by applying a Fourier magnitude projection and a real-space constraint projection
 \cite{gerchberg_optik_1972,fienup_applopt_1982,elser_josaa_2003,marchesini_rsi_2007}.
 Alternatively,
we can also frame phase retrieval as a nonlinear minimization
problem, where we minimize an error metric using a gradient-based approach. 
The gradient-based approach is flexible and can include in the forward model
a large variety of the physical phenomena related to the probing light (such as
partial coherence \cite{clark_natcomm_2012}, source fluctuations \cite{thibault_nature_2013}, 
and errors in positions
\cite{tripathi_oe_2014, dwivedi_um_2018}), 
or the detection process (such as the measurement noise \cite{thibault_njp_2012,godard_oe_2012}
and the finite size of the pixel \cite{jurling_josa_2014}).
As such, this method has been the focus of much recent literature,
leading to the development of steepest descent methods \cite{marchesini_rsi_2007,candes_itit_2015,zhang_anips_2016},
conjugate gradient methods \cite{guizar_oe_2008,marchesini_rsi_2007,wei_josaa_2017}, 
Gauss-Newton methods \cite{zhong_itci_2016}, and quasi-Newton methods \cite{li_ipi_2017}.
These algorithms have found application in the far-field ptychographic problem
not only to solve for the object alone \cite{candes_itit_2015,zhang_anips_2016,qian_ipa_2014,yan_mm_2014,yeh_ucb_2016,maiden_optica_2017} but also to additionally solve for the probe \cite{maiden_optica_2017,hesse_siamjis_2015}
as well.

Gradient-based phase retrieval methods in the literature tend to
rely on the availability of a closed-form expression for the gradient calculation.
This closed-form expression is typically obtained by writing down an 
explicit expression for the error metric to minimize,
then symbolically differentiating the error metric with respect to the individual input parameters \cite{jurling_josa_2014}. Calculating the gradient in this fashion
is laborious; a slight modification of the forward model usually requires
a complete rederivation and algorithmic reimplementation of the gradient
expressions. This becomes especially limiting if we desire to explore variations of,
or introduce new capabilities to, our basic experimental methodology.
As such, it is more than desirable to have an approach beyond symbolic differentiation  
in order to easily explore a variety of algorithms and approaches.

\textit{Automatic differentiation} \cite{griewank_siam_2008},
or \textit{algorithmic differentiation}, provides such an alternative to 
symbolic differentiation. This approach is based upon the observation that vector-valued
functions can, in general, be interpreted as composites of basic arithmetic operations
and a limited number of elementary functions (including exponentials and trigonometric
operations). Differentiation of functions can then be 
understood as a recursive
application of the chain rule of 
differentiation, wherein we repeatedly differentiate the same elementary functions
(with known derivatives), only with different input parameters. This is a mechanistic process and can hence
be performed entirely in software. 
Given a set of input numeric parameters, 
the automatic differentiation method 
computes the exact derivative by accumulating 
the numerical values of the elementary functions and their derivatives, 
without ever calculating the closed form derivative expression 
(Section \ref{sec:autodiff}).

While the field of automatic differentiation has a long and storied history \cite{griewank_siam_2008}, it is only recently that the emergence of deep 
neural network methods has driven its widespread adoption in the optimization
and machine learning communities.
Specifically, there has now arisen a need to perform gradient-based minimization to optimize state-of-the-art neural networks, which can be compositions
of thousands or even millions of individual elementary functions. 
Since calculating closed-form derivatives for these is not feasible,
automatic differentiation has become the tool of choice, thus leading to the recent rapid
adoption and advancement of such software. 
In 2014, when Jurling and Fienup first 
proposed an automatic differentiation framework for phase retrieval \cite{jurling_josa_2014},
they commented on the lack of suitable existing software packages.
Since then, we have seen the rise of multiple powerful, easy-to-use, and computationally
efficient automatic differentiation frameworks such as TensorFlow \cite{abadi_corr_2016}, PyTorch \cite{paszke_nips_2017}, and Autograd \cite{maclaurin_harvard_2016}. 
More recently, we have even seen proof-of-concept demonstrations \cite{nashed_procedia_2017, ghosh_iccp_2018} that successfully adapt these software packages to solve the phase retrieval problem.

In this paper, we first provide an overview of the \textit{reverse-mode} automatic differentiation algorithm (also referred to as the \textit{backpropagation} algorithm) for gradient calculations (Section \ref{sec:autodiff}), and then mathematically justify the application of this algorithm 
as a general framework for ptychographic phase retrieval. 
We demonstrate the numerical correctness
of the reverse-mode automatic differentiation framework
through a comparison with the popular symbolic-gradient-based 
ePIE method (Section \ref{sec:ad_ptychography}).
Finally, we demonstrate the generalizability of this framework through
successful phase retrieval for increasingly complex ptychographic 
forward models--near-field ptychography 
and 3D Bragg projection ptychography--emphasizing the flexibility 
and potential capacity of the approach for non-standard ptychography experiments
(Section \ref{sec:generalizations}). 
This shows that the reverse-mode automatic differentiation framework
allows the practitioner to update and change the forward model as 
necessary to better reflect the physics of the problem,
without prior consideration towards how to symbolically differentiate the 
error metric.  

\section{Overview of automatic differentiation}
\label{sec:autodiff}

In this section, we provide a limited overview of the automatic differentiation procedure,
focusing singularly on the \textit{reverse-mode automatic differentiation}
(or \textit{reverse-mode AD}) framework,
finally motivating the application 
of this framework to the phase retrieval problem. 
Detailed and rigorous examinations of the automatic differentiation procedure--the 
various modes and the algorithmic frameworks--are available  \cite{griewank_siam_2008}, as is
a detailed exposition of the reverse-mode AD procedure in application to the 
phase retrieval problem \cite{jurling_josa_2014}.

To demonstrate the idea of automatic differentiation, we first consider differentiable functions
$f, \phi_1, \phi_2, \phi_3:\Reals\rightarrow\Reals$ with the function composition $f=\phi_3\circ\phi_2\circ\phi_1$,
with the assumption that $\phi_1$, $\phi_2$, and $\phi_3$ are elementary functions with priorly available individual function derivatives $\dv{\phi_1}{x}$, $\dv{\phi_2}{x}$, and 
$\dv{\phi_3}{x}$ (for any $x\in\Reals$). At a given point $x=c$, 
we can compute the value of $f$ through the sequence of successive evaluations shown in Table \ref{tab:forward_backward}. We refer to this sequence of computations as the 
\textit{forward pass}.
\begin{table}[ht]
\centering
\begin{threeparttable}
\caption{Schema for reverse-mode automatic differentiation.}
\label{tab:forward_backward}
\begin{tabular}{l l c}
\textbf{Forward Pass} & \textbf{Backward Pass} & \textbf{Computational graph}\\
\hline
$\tmath{[t]&v_0 = c\\
	&v_1 = \phi_1(v_0)\\
	&v_2 = \phi_2(v_1)\\
	&v_3 = \phi_3(v_2)\\
	&f(c) = v_3}$
&
$\tmath{[t]\bar{v}_3 &=  1\\
	\bar{v}_2 &= \bar{v}_3\dv{v_3}{v_2}\\
	\bar{v}_1 &= \bar{v}_2  \dv{v_2}{v_1}\\
	\bar{v}_0 &=  \bar{v}_1 \dv{v_1}{v_0}\\
	\bar{c} &= \bar{v}_0}$
& 

\begin{tikzpicture}[node distance = 4em, 
ruler/.style={gray,->,>=stealth'},
auto, baseline=(current bounding box.north)]
\draw node [fwdc] (f0) {$v_0$};
\draw node [fwdc, right of=f0] (f1) {$v_1$};
\draw node [fwdc, right of=f1] (f2) {$v_2$};
\draw node [fwdc, right of=f2] (f3) {$v_3$};

\draw node [revc, below of=f0, node distance=4em] (r0) 
{$\bar{v}_0$};
\draw node [revc, right of=r0] (r1) 
{$\bar{v}_1$};
\draw node [revc, right of=r1] (r2) 
{$\bar{v}_2$};
\draw node [revc, right of=r2] (r3)
{$\bar{v}_3$};

\draw [line, blue] (f0) -- (f1) node
[midway, above, align=center, text=blue] {$\phi_1$};
\draw [line, blue] (f1) -- (f2) node
[midway, above, align=center, text=blue] {$\phi_2$};
\draw [line, blue] (f2) -- (f3) node
[midway, above, align=center, text=blue] {$\phi_3$};

\draw [line, red, dashed] (r3) -- (r2) node
[midway, above, align=center, text=red]{$\dv{v_3}{v_2}$};
\draw [line, red, dashed] (r2) -- (r1) node
[midway, above, align=center, text=red] {$\dv{v_2}{v_1}$};
\draw [line, red, dashed] (r1) -- (r0) node
[midway, above, align=center, text=red]{$\dv{v_1}{v_0}$};

\draw [line, blue] ($(f0.north)+(-0.25,0.25)$) -- ++(4.75,0) node(f3.west) 
[midway, above, align=center, text=blue] 
{Forward pass};
\draw [line, dashed, red] ($(r3.south)+(0.25,-0.25)$) -- ++(-4.75,0) node(r0.east) 
[midway, below, align=center, text=red] 
{Backward pass};
\path [line, dotted] (f0) -- (r0) node {};
\path [line, dotted] (f1) -- (r1) node {};
\path [line, dotted] (f2) -- (r2) node {};
\end{tikzpicture}\\
\hline
\end{tabular}
\begin{tablenotes}
	\small
	\item The gradient is evaluated
		through an accumulation of intermediate values calculated for the individual
		elementary functions and their derivatives.
\end{tablenotes}
\end{threeparttable}
\end{table}

In the evaluation trace shown in Table \ref{tab:forward_backward},
 we follow the notation in the literature \cite{griewank_siam_2008,baydin_jmlr_2015,hoffmann_na_2016} 
and index the variables stored in the memory as $v_i$, with 
$i\leq 0$ for the input variables, and $i>0$ for the intermediate computed variables.
To calculate the derivative of $f$ at $x=c$, we use the chain rule of 
differentiation:
\begin{align*}
	\dv{f}{x}\biggr|_{x=c} &= \dv{\phi_3}{x}\biggr|_{x=\phi_2(\phi_1(c))}
									\cdot\dv{\phi_2}{x}\biggr|_{x=\phi_1(c)}
									\cdot\dv{\phi_1}{x}\biggr|_{x=c}.
\end{align*}
To evaluate this derivative, we perform a backward pass
(\textit{i.e.}, a reversed sequence of computations)
during which we associate each 
intermediate variable from the forward pass, $v_i$, with a new adjoint
variable $\bar{v}_i = \dv{f}{v_i}$; we then evaluate $\bar{v}_i$ from $i=3$ to $i=0$.
This is shown in Table \ref{tab:forward_backward}.
Noting that 
\begin{equation}
 \dv{v_{i+1}}{v_i} = \dv{\phi_{i+1}}{x}\biggr|_{x=v_i},
\end{equation}
we can see that each step in the backward pass requires
as input not only the derivative values calculated in the previous steps, 
but also the intermediate variables calculated during the forward pass.
This is illustrated in the computational graph in Table \ref{tab:forward_backward}.
Thus, to calculate the final derivative value, we need to: 
\begin{enumerate}
	\item identify the derivatives of the elementary functions $\phi_1$, $\phi_2$, and $\phi_3$,
	\item perform a forward pass evaluation of the function and store the intermediate values calculated, and
	\item perform a backward pass to accumulate the final derivative.
\end{enumerate}
This reverse-mode automatic differentiation scheme calculates the exact derivative value
(up to floating point errors) without relying on the closed form expression
of the derivative; instead, it relies only on the structure of the computational graph. 

From Table \ref{tab:forward_backward},
we can see that for the reverse-mode AD gradient calculation, 
the values of the intermediate variables calculated during the forward pass are shared with the backward pass; 
each elementary expression is only computed once but reused multiple times. 
This ensures that the gradient calculation is very efficient computationally. 
In fact, for functions of the form $f:\Reals^n\rightarrow\Reals$, if the function evaluation (forward pass) requires 
\begin{align*}
	\text{ops}(f) = N
\end{align*}
floating point operations, then the number of floating point operations required for the
gradient evaluation (backward pass) is always given by 
\begin{align*}
\text{ops}(\grad f) = k\cdot \text{ops}(f) = k N,
\quad\text{with }0<k<6\text{ a constant},
\end{align*}
such that $k$ typically satisfies $k\in[2,3]$, no matter the value of $n$ \cite{griewank_siam_2008,baydin_jmlr_2015}.
In other words, for reverse-mode AD, barring memory limitations,
the time required to calculate the gradient
$\grad f$ is \textit{always} within an order of magnitude of the time required to calculate the function value itself. 
This is known as the \textit{cheap gradient principle}.

As such, the reverse-mode AD procedure is ideally suited to provide gradient-based
iterations aiming at solving of optimization problems. A quintessential case in point
is machine learning with neural networks--the `training step' in large neural
networks boils down to the numerical optimization of an error metric
involving a large number of functional compositions and up to millions of 
input parameters. In such a situation, the derivation of the closed-form
gradient with pen and paper is simply out of reach.
Thus, the implementation of gradient descent procedures that use the 
AD framework has been key to the recent meteoric rise in machine
learning applications. 

In contrast to the neural network case, 
a typical error metric for the phase retrieval problem only
involves a limited number of functional compositions, and the closed form gradients
can be calculated manually. This has been demonstrated in prior phase retrieval literature:
the Wirtinger flow method and its variants \cite{candes_itit_2015,zhang_anips_2016,bian_oe_2015},
the PIE family of methods \cite{maiden_optica_2017}, 
and numerous other methods proposed in the literature 
\cite{fienup_applopt_1982,marchesini_rsi_2007,wen_ip_2012,godard_oe_2012,qian_ipa_2014}
are examples of gradient-based descent procedures that rely on such closed-form gradient expressions. However, the impressively flexible AD framework not only simplifies these selfsame gradient calculations, but also allows us to modify the forward model; this empowers us to address the full range of problems that are related to phase 
retrieval. Consequently, we expect that the AD framework should also greatly benefit the
phase retrieval community.

As we demonstrate in Section \ref{sec:ad_ptychography} for the ptychographic problem,
the error metric for the phase retrieval problem is a scalar-valued multivariate objective function of complex variables,  $f:\Complexes^n\rightarrow\Reals$ \cite{fienup_josa_1982,marchesini_rsi_2007}. To minimize such a function using a gradient descent approach,
we adopt the Wirtinger gradient descent
formalism \cite{brandwood_iproc_1983,kreutz_arxiv_2009,sorber_siam_2012}. For some $z\in\Complexes^n$,
the Wirtinger gradient operator is defined as
\begin{align}
	\grad{f(z)} = \pdv{f}{z^*} = \frac{1}{2}\left(\pdv{f}{\Rel{z}} 
	+ i\pdv{f}{\Img{z}}\right),
	\label{wirtinger}
\end{align}
where $i=\sqrt{-1}$, $z^*$ is the element-wise complex conjugate of $z$, $\Rel{z}$ and $\Img{z}$ 
are respectively the real and imaginary parts of $z$, and $\pdv{f}{\Rel{z}}$
and $\pdv{f}{\Img{z}}$ the componentwise partial derivatives with respect to the 
real and imaginary parts of $z$.  
Since the partial derivatives $\pdv{f}{\Rel{z}}$
and $\pdv{f}{\Img{z}}$ are both individually real-valued, 
we can calculate them by separately using the reverse-mode AD framework
in the same fashion as shown in Table \ref{tab:forward_backward}.
This ensures that the gradient calculation procedure is very efficient, 
with the time cost once again comparable to that for the objective function itself.

\section{Validation: Far-field transmission ptychography}
\label{sec:ad_ptychography}
In this section, we first establish a forward model for far-field transmission ptychography.
We then use reverse-mode AD to set up a gradient descent procedure  that 
is, by construction, equivalent to the ePIE reconstruction method  \cite{maiden_optica_2017}
which uses a closed-form gradient expression. We compare these frameworks numerically
and establish that the automatic differentiation framework calculates 
gradient values that are identical to those calculated via the
closed-form gradient expressions.
This comparison to the well-known ePIE method serves to establish the validity
of the AD framework for phase retrieval. The ePIE method, however, cannot be used
out-of-the-box for object reconstruction once we modify the ptychographic forward model--as such, it is not well-suited for use with the AD framework. 
Instead, we demonstrate that
we can use the flexible and easy-to-use state-of-the-art accelerated adaptive gradient 
descent algorithms (like the Adaptive Moment Estimation or \textit{Adam} algorithm \cite{kingma_corr_2014})
that are commonly available out-of-the-box with AD software to efficiently solve the ptychographic
reconstruction problem.

\subsection{Forward model for far-field transmission ptychography}
\label{sec:forward_model_farfield}

In far-field transmission ptychography, an unknown object is illuminated with a coherent beam,
called the probe, which is localized to a small area on the object. 
The intensity of the wavefront transmitted through the object is then 
measured at the far field by a pixel-array detector. 
The beam is used to raster scan the object in a grid of $J$ spatially
overlapping illumination spots, generating a set of $J$ diffraction patterns at the detector plane. 
The forward model for far-field ptychography consists of the following steps:
\begin{enumerate}
	\item The complex valued object transmission function $\OO$ is approximated by a total of $N$ pixels in the object plane. 
	For the illumination position $\rr_j$, a shift operator $\SSj$, with an $M\times N$ matrix representation, 
	extracts the $M$ object pixels illuminated by the probe beam $\PP$ containing $M$ pixels in the object plane. 
	The thus transmitted wave function is represented by the exit wave 
	$\psi_j\in\Complexes^M$:
	\begin{align}
	\psi_j = \PP \odot (\SSj \OO) &\quad \mathrm{for}\;j = 1, 2, ..., J;
	\label{fwdmodel_exitwave}
	\end{align}
	where $\odot$ is the element-wise Hadamard product operator.
	\item Each transmitted exit wave $\psi_j$ propagates to the far field detector plane, where the detector then records a real-valued intensity pattern containing $M$ pixels.
	If there is no noise fluctuation, the expected intensity of this $j$th wave-field at
	the detector plane reads
	\begin{align}
	h_j = |\boldsymbol{\mathcal{F}}\psi_j|^2 + \nu_j
	\label{expected_forward_model}
	\end{align}
	where $\nu_j$ is the expected level of background events, $\FF$ 
	is the matrix representation of the two-dimensional digital Fourier transform, and $|\cdot|$ 
	extracts the modulus element-wise for a complex-valued vector. 
	
	\item The model in Eq. \eqref{expected_forward_model} describes the behavior of the detected intensity 
	in average. During an experiment, each diffraction pattern is subject to random 
	fluctuations produced by the instrumental and the Poisson counting fluctuations (shot-noise). 
	If one further assumes that the instrumental (thermal) noise is negligible, a Gaussian additive 
	perturbation is usually accurate enough to describe how the counting fluctuation plagues the 
	square-root of the measured intensity (see, for instance, \cite{godard_oe_2012} and references therein):
	\begin{align}
	y_j^{1/2} = h_j^{1/2} + \varepsilon_j
	\label{noise_forward_model}
	\end{align}
	where $\varepsilon_j$ is a centered Gaussian random vector. 
\end{enumerate}
The above relations are the forward (observation) model that predicts how 
the observations $\{y_j^{1/2}\}_{j=1}^J$ behave when the sample and the probe 
are jointly given. The inversion/reconstruction step simply aims at retrieving 
both these quantities from the observations. For that purpose, the maximum likelihood
estimator defines the solution of this reconstruction step via a minimization
problem that can be easily derived from the fluctuation model of
Eq.~\eqref{noise_forward_model}.      
In our case it leads to 
\begin{align}
(\OO^\star,\PP^\star) \in \text{argmin}_{_\OO,_\mathcal{P}} g(\OO,\PP)
\label{maximum_likelihood_estimator}
\end{align}
where $g$ is a separable fitting-function that reads 
\begin{equation}
g = \sum_{j=1}^J g_j \quad \text{with} \quad g_j (\mathcal{O},\mathcal{P}) := || y_j^{1/2} - h_j^{1/2}(\mathcal{O},\mathcal{P}) ||^2
\label{fitting_function}
\end{equation}
where $\norm{\cdot}$ denotes the usual Euclidian norm in $\mathbb{C}^N$, and $h_j^{1/2}$ and $y_j^{1/2}$ denote the componentwise 
square roots of the vectors $h_j$ and $y_j$ respectively. 
While this fitting-function has been used in the phase retrieval literature for decades, there exist some alternative noise models that would in turn provide alternative functionals 
to minimize \cite{godard_oe_2012,guizar_oe_2008}. In any case, the optimization step derived from the noise model involves 
a large-scale phase retrieval problem that is rather challenging. Phase retrieval problems 
are NP-hard problems because of their inherent non-convex structure and the search for good 
and computationally efficient heuristics is still an open problem. The overlapping between 
successive scanning probe is however recognized as a key factor that helps in preventing 
stagnation for standard, derivative-based iterations. Thanks to AD, these derivatives can
be efficiently computed and further used by one of the iterative solvers that will be discussed
in the next section.     

\subsection{Iterative ptychographic reconstruction with reverse-mode AD: ePIE and Adam}

We first consider the \textit{extended Ptychographical Iterative Engine} or ePIE algorithm \cite{maiden_ultramic_2009}, which can be summarized in a generalized form in Algorithm \ref{alg:epie_farfield}.
\begin{algorithm}
	\caption{Generalized ePIE gradient descent iteration}
	\begin{algorithmic}[1]
		\Require Probe and object initial guesses $\PP^0$ and $\OO^0$.
		\Require Minibatch size $b$. For the original ePIE procedure, $b=1$.
		\State Randomly shuffle the probe positions $\mathfrak{J} =\{1,2,...,J\}\xrightarrow{\text{shuffle}}\mathfrak{J}'$.
		\For {$k=1$ to $J/b$}
		\State With the probe positions indexed as $j\in\mathfrak{J}'$, calculate the partial derivatives:
		\begin{align}
		\delO g^k := \sum_{j=b(k-1)+1}^{bk} \delO g(\OO^{k-1}_j)\quad\text{and}\quad
		\delP g^k := \sum_{j=b(k-1)+1}^{bk} \delP g (\PP^{k-1}_j)
		\label{minibatch_grads}
		\end{align}
		\State Calculate the step sizes
		\begin{align}
		\alpha_\sOO^k :=  1/\maxnorm{\norm{\PP^{k-1}}^2}\quad\text{and}\quad
		\alpha_\sPP^k :=  1/\maxnorm{\sum_{j=b(k-1)+1}^{bk} \norm{\SSj^*\OO^{k-1}}^2}.
		\label{epie_lipschitz}
		\end{align}
		where $\maxnorm{\cdot}$ denotes the maximum element of the enclosed vector.
		\State Set
		\begin{align}
		\OO^k = \OO^{k-1} - \alpha_\sOO^k \delO g^k\quad\text{and}\quad
		\PP^k = \PP^{k-1} - \alpha_\sPP^k \delP g^k
		\label{epie_update}
		\end{align}
		\EndFor
	\end{algorithmic}
	\label{alg:epie_farfield}
\end{algorithm}

With $b=1$, Algorithm \ref{alg:epie_farfield} exhibits one iteration of the ePIE procedure, during which it uses the information in 
single diffraction patterns to cyclically update the current probe and object estimates.
The core (or ``Engine'') of this update is the correction (Eq. \eqref{epie_update}) computed from the considered probe position $j\in\mathfrak{J}'$ using the derivative of
$g$ with respect to the sample $\OO$ and the probe $\PP$. In its original formulation, the 
ePIE procedure uses the well-known closed form expression of these derivatives 
(see Eqs. (6) and (7) in \cite{maiden_ultramic_2009}) to compute these updates.

Alternatively, we can use the reverse-mode AD procedure for the numerical computation of these derivatives by relying on the Wirtinger formalism (Eq. \eqref{wirtinger}) for complex derivatives. The derivative with respect to the real and imaginary parts of the complex 
variable are evaluated separately, and they are then accumulated according to the equations
\begin{equation}
\delO g^k := \frac{1}{2}\left(\pdv{g}
{\Rel{\OO}}\Bigg\rvert_{\OO^{k-1}}+ i\pdv{g}{\Img{\OO}}\Bigg\rvert_{\OO=\OO^{k-1}}\right)
\label{Derivative_wrt_O}
\end{equation}
and
\begin{equation}
\delP  g^k := \frac{1}{2}\left( \pdv{g}{\Rel{\PP}}\Bigg\rvert_{\PP=\PP^{k-1}} + i\pdv{g}{\Img{\PP}}\Bigg\rvert_{\PP=\PP^{k-1}}\right).
\label{Derivative_wrt_P}
\end{equation}
The AD version of ePIE is mathematically 
equivalent to the original ePIE \cite{maiden_ultramic_2009}; thus, both the algorithms should generate the same sequence of iterates. 
We refer to the AD version of the ePIE solver as AD-ePIE.

\begin{figure}[ht]
	\centering
	\begin{tikzpicture}[node distance = 4em, 
	ruler/.style={gray,->,>=stealth'},
	auto]
	\draw node [var] (f0) {Object($\OO$)};
	
	\draw node [fwd, right of=f0, node distance=2.75cm, fill=green!60!black!20] (f1) 
	{$\psi_j = \PP\odot(\SSj\OO)$};
	
	\draw node [fwd, right of=f1, node distance=3.0cm, fill=green!60!black!20] (f2)
	{$h_j = \norm{\FF\psi_j}^2 + \nu_j$};
	
	\draw node [fwd, right of=f2, node distance=3.55cm, fill=red!80!black!20] (f3)
	{$\tmath{g(\OO) = \sumj  \norm{\sqrt{y_j}- \sqrt{h_j}}^2}$};
	
	\draw node [var, above of=f1, node distance=2cm] (f4)
	{Probe($\PP$)};
	
	\draw node [var, above of=f3, node distance=2cm, fill=red!30!black!30] (f5)
	{Intensities($y_j$)};
	
	\draw node [above of=f2, color=blue, align=center, node distance=1cm] 
	{Far-field\\detection};
	
	
	\draw node [below of=f1, color=red]
	{$\tmath{\pdv{g}{\OO^*} = \pdv{\psi_j}{\OO^*}\pdv{g}{\psi_j}}$};
	
	\draw node [below of=f2, color=red]
	{$\tmath{\pdv{g}{\psi_j} = \pdv{y_j}{\psi_j}\pdv{g}{h_j}}$};
	
	\draw node [below of=f3, color=red]
	{$\tmath{\pdv{g}{y_j}}$};
	
	\path [line, blue] (f0) -- (f1);
	\path [line, blue] (f1) -- (f2);
	\path [line, blue] (f2) -- (f3);
	\path [line, blue] (f4) -- (f1) node [midway, left, align=center, text=blue]
	{Illumination\\position $j$};
	\path [line, blue] (f5) -- (f3) node [midway, left, align=center, text=blue]
	{Error\\function};
	\path [line, orange, dashed] (f3.south) to [out=210, in=-30] (f2.south);
	\path [line, orange, dashed] (f2.south) to [out=210, in=-30] (f1.south);
	\path [line, orange, dashed] (f1.south) to [out=210, in=-30] (f0.south);
	\end{tikzpicture}
	\caption{Simplified representation of the forward and backward passes for the 
		object update for far-field ptychography.
		The solid blue arrows indicate the forward pass direction. 
		The dashed orange arrows indicate the backward pass direction.}
	\label{fig:farfield_graph}
\end{figure}
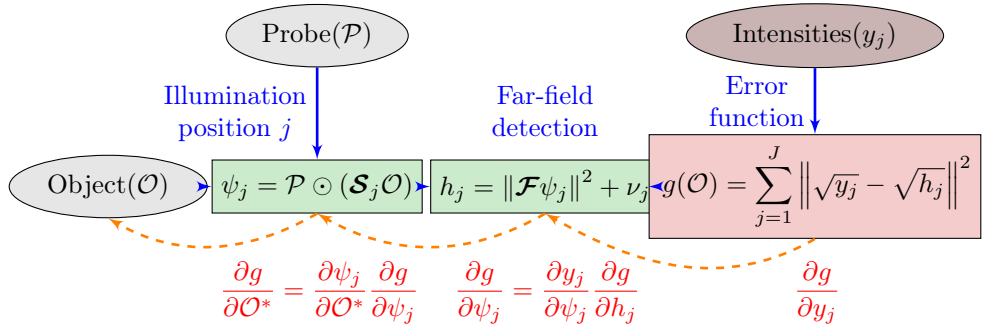

In \Cref{fig:farfield_graph} we present a simplified representation
of the computational graph structure that generates the AD-ePIE iterates (see also Table \ref{tab:forward_backward}). 
We can see from the figure that the modular reverse-mode AD framework is 
agnostic as to the choice of the error
metric, the update scheme, and even to the forward model itself; 
it generalizes straightforwardly to any variations in these.
We demonstrate this in Section \ref{sec:generalizations},
where we vary both the forward model and the update scheme.

A natural extension to the ePIE/AD-ePIE algorithm is to use several
probe positions to compute a single update: that is, to increase  the minibatch size $b \geq 1$ in Algorithm \ref{alg:epie_farfield}. 
In this perspective, 
Algorithm 1 belongs to a larger family of iterates that takes advantage of the natural partitioning of the dataset \cite{godard_oe_2012}. Similar or identical optimization strategies are indeed well known under different names in 
several communities, such as `ordered-subset’ in image processing \cite{ahn_ieee_2003}, `incremental gradient’
\cite[Chapter\ 1]{bertsekas_athena_1999} in the optimization literature, or `stochastic gradient’ for neural-network 
learning algorithms \cite{lecun_springer_2012}.

A salient feature of ePIE/AD-ePIE is that the chosen step sizes $\alpha_\sOO^k$ 
and $\alpha_\sPP^k$ (Eq. \eqref{epie_lipschitz}) for the object and probe updates respectively equate to the inverse of the \textit{Lipschitz constant} of the partial gradients $\delO g^k$ and $\delP g^k$ \cite{hesse_siamjis_2015}.
This choice of step sizes is well-known in the optimization literature, 
and is particularly useful in a batch gradient descent setting where the
derivatives in the update (Eq. \eqref{epie_update}) are calculated using \textit{all} 
available probe positions at once (setting $b=J$).  As an example, with
a steepest-descent 
iteration algorithm \cite{bertsekas_athena_1999} these step sizes would then ensure global 
convergence (toward a local minimizer). 
However, the Lipschitz constants are
not usually known \textit{a priori}; they generally have to be carefully derived
using the closed-form gradient expressions \cite{hesse_siamjis_2015}.
Changes to the forward model, or the inclusion of additional terms in the error metric (e.g. regularizers),
can thus require a re-deriving of both the closed form gradient expression and the Lipschitz constant.
As such, using the Lipschitz constants
to calculate the step sizes would negate the flexibility that is the hallmark of the AD procedure.

To circumvent this limitation and enable phase retrieval within the AD framework, we can substitute the ePIE/AD-ePIE method with 
a choice of state-of-the-art adaptive gradient descent algorithms that are widely used in the
machine learning literature \cite{ruder_arxiv_2016}, such as Momentum, Nesterov Momentum, 
Adagrad, Adadelta, RMSProp, or Adam. The performance of these exotic methods 
specifically in phase retrieval applications is a new but promising area of research, 
with the recent literature \cite{maiden_optica_2017,pauwels_itsp_2018,xu_arxiv_2018}
demonstrating that momentum-based accelerated gradient descent methods converge to a solution
faster than standard gradient descent methods. In this work, we use the Adam (Adaptive Moment Estimation) 
gradient descent procedure \cite{kingma_corr_2014} that uses a momentum-like acceleration and, crucially, does not rely on the 
Lipschitz constant for the gradient descent. As such, the Adam method is robust to changes in the error 
metric and therefore well-suited to phase retrieval applications.

The Adam optimization method is available out-of-the-box in the commonly used AD toolsets
TensorFlow \cite{abadi_corr_2016} and PyTorch \cite{paszke_nips_2017},
so that it can be used directly
without first implementing the underlying algorithm. Nevertheless, for the sake of clarity, 
we present in Appendix \ref{sec:adam_alg}
the parameter initialization step and the variable update step
that make up the Adam optimizer for the object variable (the probe updates are
similarly calculated). 
To solve the ptychography problem, we incorporate the Adam optimizer 
in a fashion identical to Algorithm \ref{alg:epie_farfield}, but 
with steps 4 and 5 substituted with the update computed from the Adam optimizer.
In Section \ref{sec:results_farfield}, we provide a minimal comparison of performance of the Adam method to 
the ePIE/AD-ePIE method by iterating through the individual probe positions and calculating the 
object and probe updates separately per probe position. 
In Section \ref{sec:hyperparameters}, we present the search strategy used for the
	choice of the Adam \textit{hyperparameters} (viz., the initial probe update step size, the initial object update step size, and the minibatch size) adopted in this work, and provide heuristic guidelines for 
	the choice of hyperparameters to achieve computationally efficient ptychographic reconstruction. Beyond our heuristic approach, a more detailed evaluation of the
application of the Adam algorithm for phase retrieval is an important research question but 
is beyond the scope of this paper.

\subsection{Numerical results}
\label{sec:results_farfield}

For a numerical validation of the reverse-mode AD procedure, we simulated a far-field 
transmission ptychography experiment with an incident x-ray of energy $8.7$ keV.
We used the `Mandrill' and `Cameraman' images, each $128\times128$ pixels in size,
as the test object magnitude and phase respectively, 
and embedded the test object at the center of a simulation box of size $190\times190$ pixels (with tight support).
We illuminated the test object 
with a complex-valued probe function approximated using an array of size $64\times 64$ pixels and with a total
integrated intensity of $10^6$ photons at the object plane. The probe function was obtained by 
propagating the exit wave from a square aperture
of width $7\,\mu$m to a distance of $\zeta=15$ cm so that it contained diffraction fringes
characterized by $\sqrt{\lambda \zeta} = 4.6\,\mu$m.  
The raster grid for the object scan was obtained by translating the probe latitudinally
and longitudinally in steps of 3.5 $\mu$m (6 pixels). The real-space pixel grid was obtained
by assuming exact Nyquist sampling with respect to a detector with a pixel pitch of 55 $\mu$m placed at a distance of $14.6$ m from the object. 
We thereby generated 484 far-field diffraction patterns at the detector position,
then incorporated Poisson noise into these diffraction patterns to get the simulated 
far-field data points.
\begin{figure}[th]
	\centering
	\begin{tabular}{ccc}
	\includegraphics[width=0.3\linewidth]{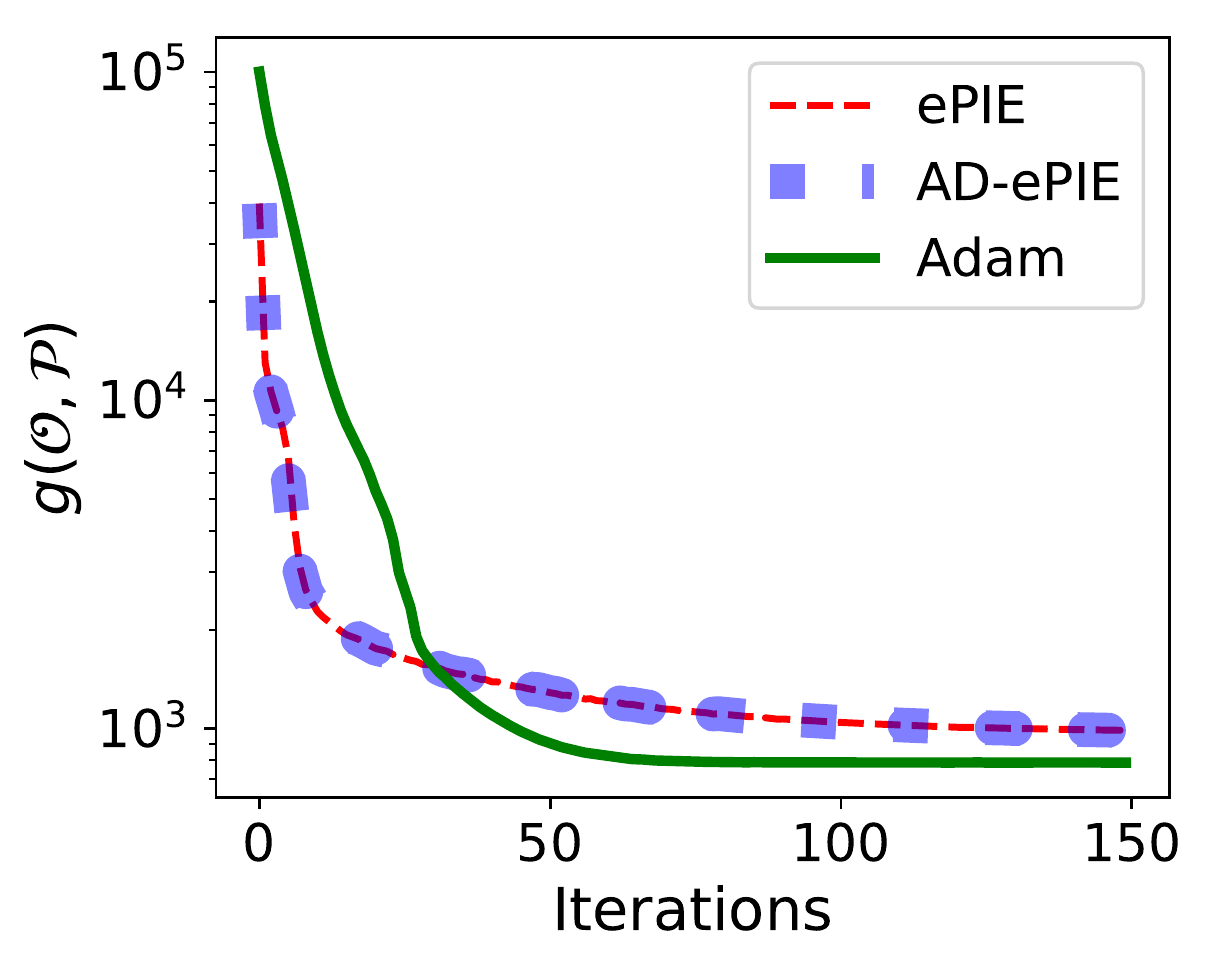} &
	\includegraphics[width=0.3\linewidth]{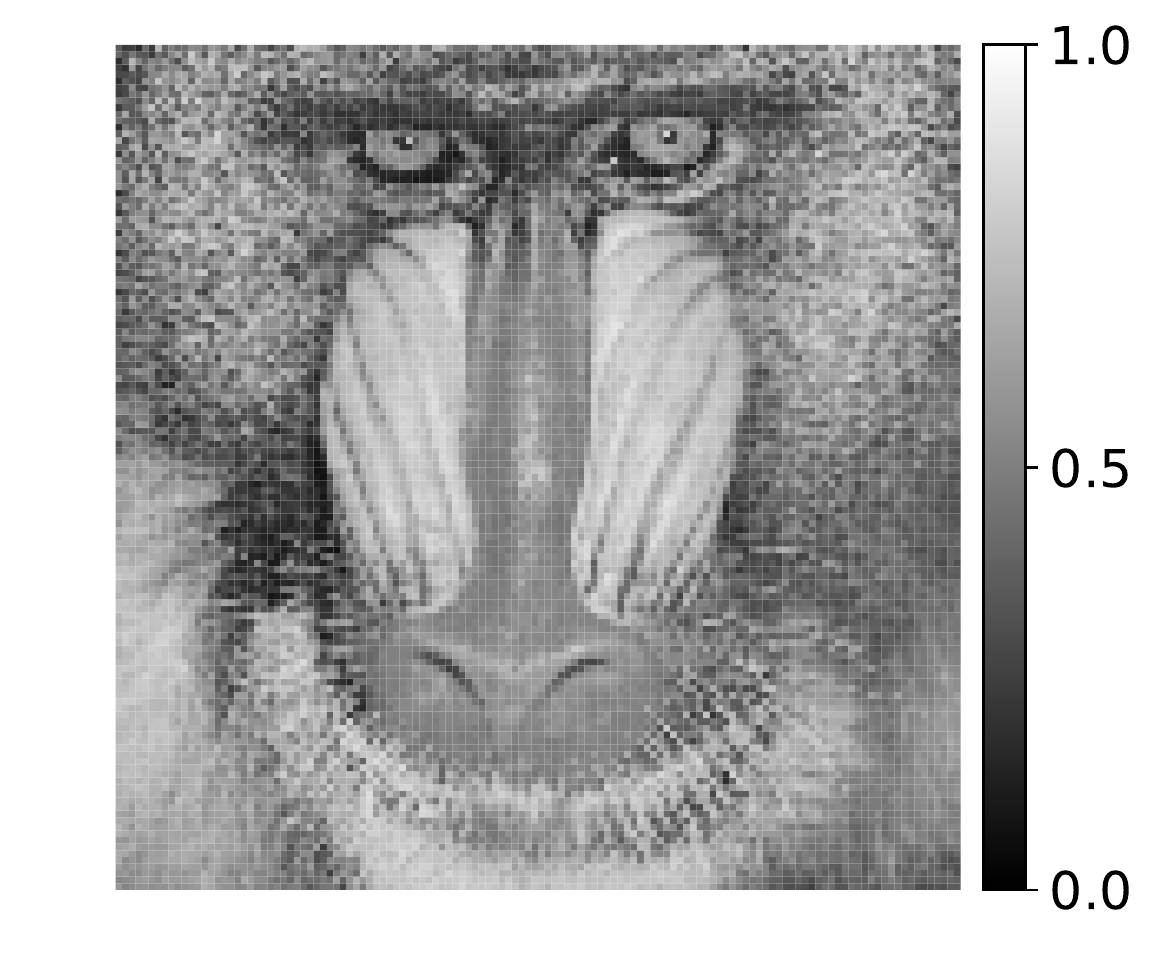} &
	\includegraphics[width=0.31\linewidth]{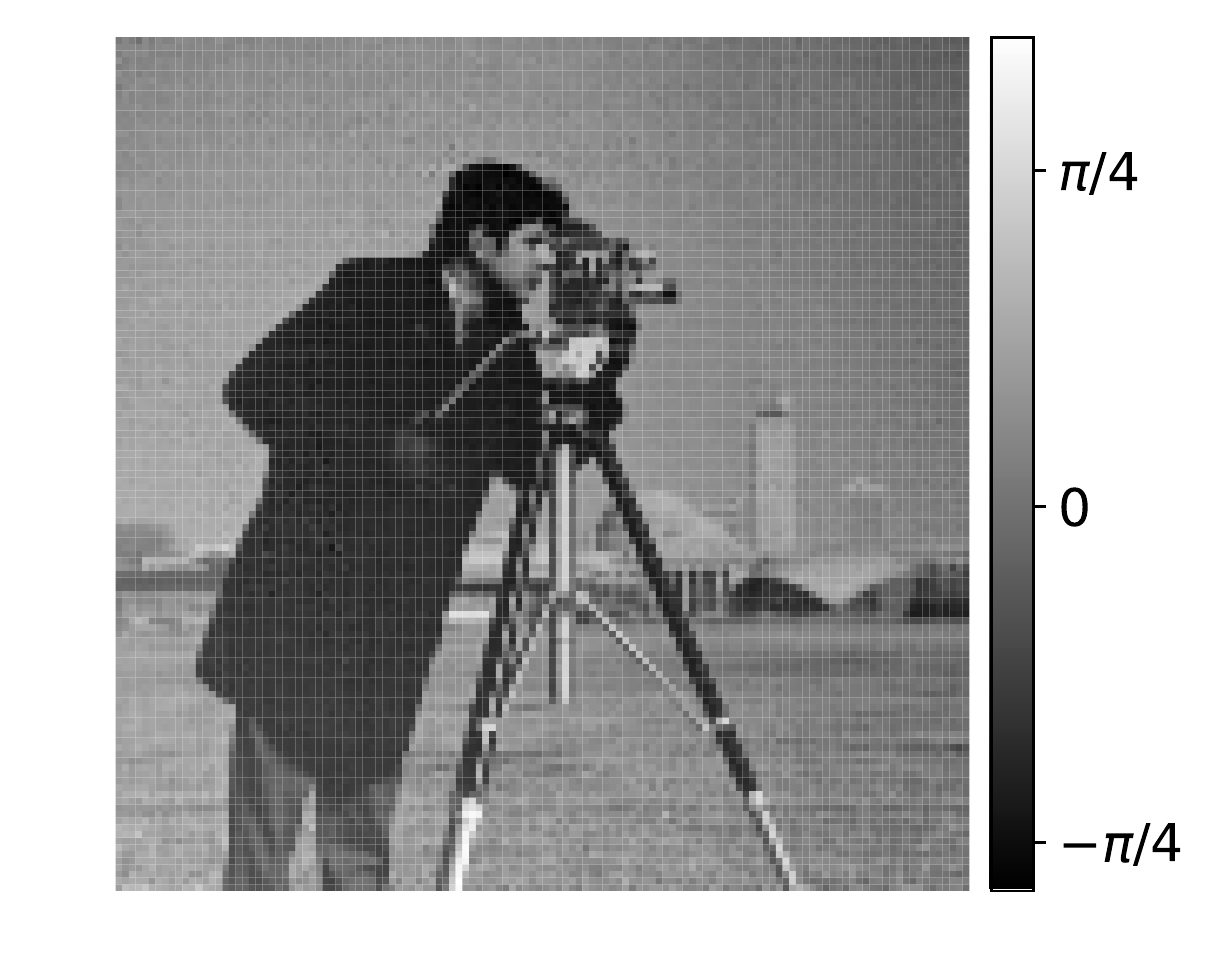} \\
	(a) & (b) & (c) \\
	\includegraphics[width=0.3\linewidth]{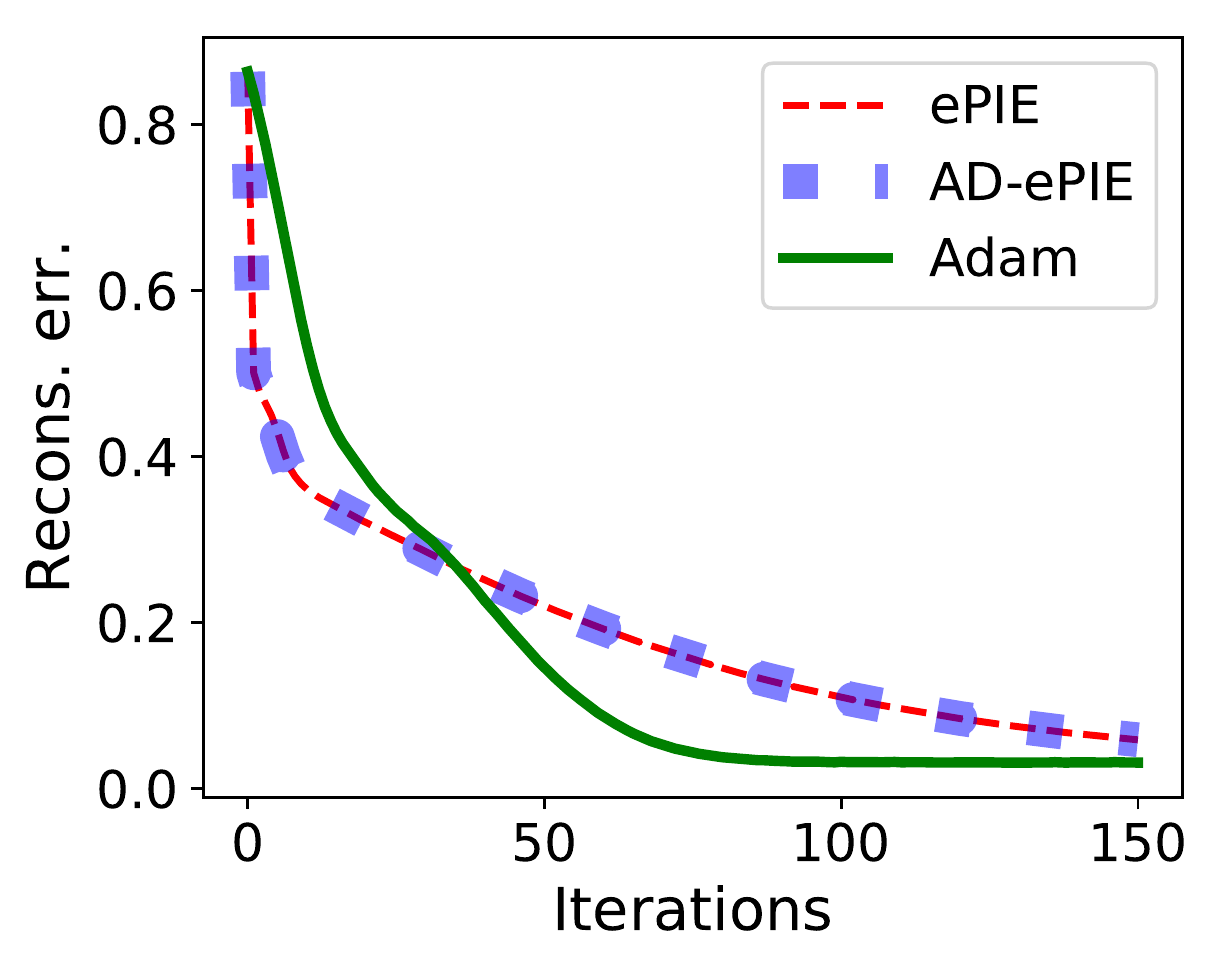} &
	\includegraphics[width=0.3\linewidth]{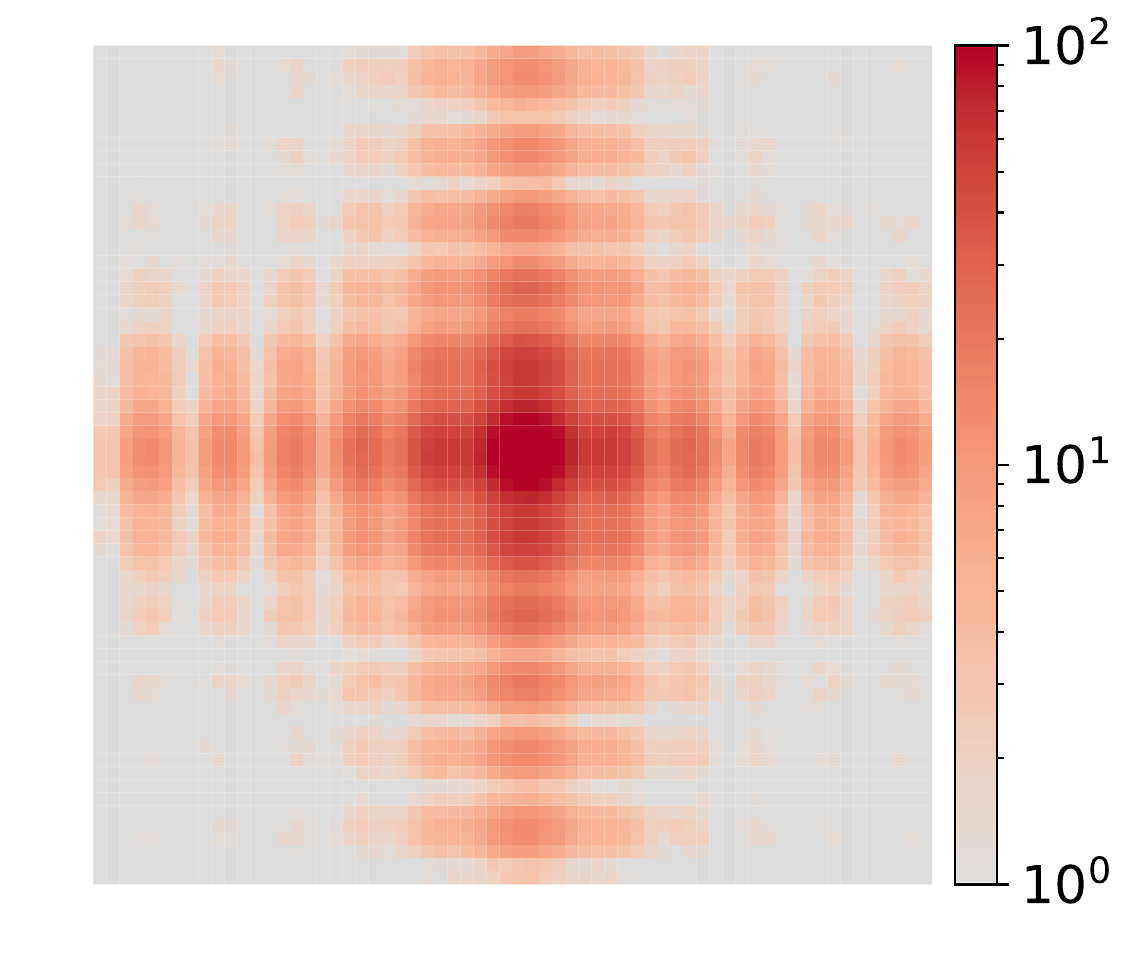} &
	\includegraphics[width=0.3\linewidth]{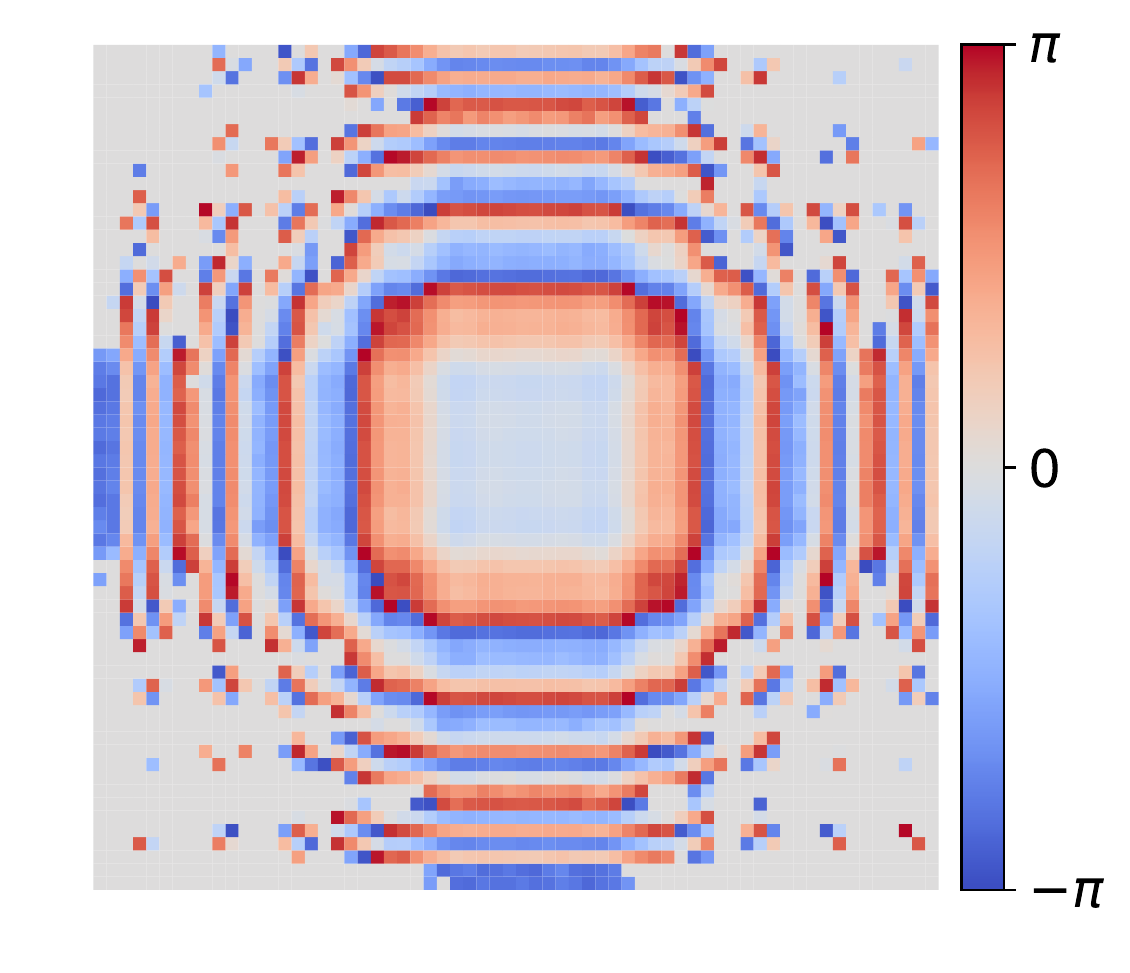} \\
	(d) & (e) & (f)
\end{tabular}
	\caption{The value of the (a) error metric $g(\OO, \PP)$ and (d) the normalized reconstruction
		error for the object for the ePIE method, the AD-ePIE method,
		and the Adam method for the far-field ptychography experiment. Adam reconstructions for the (b) object magnitude, (c) object phase,
		(e) probe magnitude, and (f) probe phase. The normalized root-mean-squared reconstruction error (NRMSE)
		was 0.03 for the object and 0.02 for the probe.}
	\label{fig:farfield_results}
\end{figure}

Our implementation of the AD-based ptychographic reconstruction framework follows previous work \cite{nashed_procedia_2017,ghosh_iccp_2018} and uses the
TensorFlow \cite{abadi_corr_2016} machine learning library for the gradient calculation.
For the reconstruction, we used a $7\,\mu$m (12 pixels) square aperture as the initial probe guess,
and a random complex array as the initial object guess. Using the same
starting parameters, we performed 150 iterations through the data set 
(with the probe value held constant for the first iteration) for the ePIE, AD-ePIE, and the Adam methods,
with the same randomized sequence of diffraction patterns for all three algorithms. 
For the Adam method, we used initial update step sizes $0.1$ and $0.001$ for the probe and object respectively.

As we demonstrate in \Cref{fig:farfield_results}, the AD-ePIE method
followed the same reconstruction path as the ePIE method for both the error metric $g(\OO, \PP)$, 
and the normalized object reconstruction error calculated per pixel from the ground truth \cite{guizar_ol_2008}. 
This demonstrates that the reverse-mode AD framework described in this paper calculates gradient 
values numerically identical to those calculated using the closed form symbolic derivatives \cite{maiden_ultramic_2009}. Additionally, reconstruction using the Adam algorithm also converges to the same final probe and object structures as the ePIE algorithm: the final Adam object and probe structures only differ by $5\%$ and $4\%$ respectively from the corresponding final ePIE structures.

The reconstruction algorithms implemented in this work do not use any advanced domain decomposition techniques to parallelize the ptychographic reconstruction \cite{nashed_optexp_2014,nashed_procedia_2017}; they iterate sequentially through minibatches of diffraction patterns. When these algorithms were implemented to run on a single 3.00 GHz Intel Xeon processor with a minibatch $b=1$, the runtime for each minibatch update for the forward model and the ePIE algorithm (implemented with the \textit{numpy} library in Python) were found to be $\approx1$ ms, and $\approx1.2$ ms respectively. The corresponding runtime for the AD-ePIE and Adam algorithms (implemented with TensorFlow) were found to be $\approx4$ ms, and $\approx5.8$ ms respectively. Disregarding backend discrepancies, this is in agreement with the expected computational costs described in Section \ref{sec:autodiff}---the ePIE algorithm implements the symbolic gradient expressions directly and has a runtime comparable to the forward model itself. The AD-ePIE method has a runtime that is within a small scaling factor of that for the forward model, and the Adam algorithm requires some additional computation for the necessary moment updates (Algorithm \ref{alg:adam}). In practice, the Adam algorithm converges to the final solution in 80 iterations ($\approx 225$ s) while the ePIE algorithm requires 150 iterations ($\approx 87$ s) to achieve convergence. Once we use larger minibatch sizes, however, as we demonstrate in Section \ref{sec:hyperparameters}, the Adam algorithm can converge to a solution much faster than the basic ePIE approach.

\subsection{Choosing Adam hyperparameters for efficient ptychographic reconstruction}
\label{sec:hyperparameters}

In this work, for reconstructions with the Adam algorithm (Algorithms \ref{alg:epie_farfield} and \ref{alg:adam}), we manually supply three key hyperparameters: the minibatch size $b$, and the initial Adam object and probe update step sizes $\alpha_\sOO^A$ and $\alpha_\sPP^A$. For optimal performance of these 
reconstruction algorithms, we need to choose hyperparameter values that simultaneously optimize the hardware utilization, the time cost per minibatch update, and the total number of minibatch updates required to converge to a solution---this makes for a difficult research problem \cite{bergstra_jmlr_2012}. As such, in this work, we take a primarily heuristic two-step approach: 1) we pick a minibatch size that optimizes the hardware utilization and time cost per minibatch update, and then, with this number size fixed, 2) we perform a gridsearch to identify the values for $\alpha_\sOO^A$ and $\alpha_\sPP^A$ that reduce the number of iterations required for the reconstruction.

Ptychographic reconstruction with $b=1$, as used in Section \ref{sec:results_farfield}, has optimally low time cost per minibatch update, but is not suitable for implementation on parallel computing hardware. In fact, when implemented for use in a nVIDIA K40 GPU, this reconstruction procedure has minimal GPU utilization ($<10\%$), and shows no noticeable improvement over the CPU version in the time cost per minibatch update. This suggests the use of larger minibatches for optimal GPU utilization. As we increase the minibatch size, however, the number of intermediate values stored in the memory also increases proportionally.	As such, we need to choose a minibatch size that ensures that the resulting computational graph, including all the intermediate values calculated in the forward and backward passes, fits within the GPU memory. For the ptychographic dataset simulated in Section \ref{sec:results_farfield}, we can set $b=100$; this choice combines much greater GPU utilization ($>50\%$) with minimal time cost per minibatch update ($\approx 5.2$ ms), and also maintains stochasticity in the gradient directions computed per update.

With the minibatch size fixed, we can then perform a gridsearch to identify the $\alpha_\sOO^A$ and $\alpha_\sPP^A$ values that lead to fast ptychographic reconstruction. While there exist sophisticated methods to modify these parameters within the descent procedure \cite{kingma_corr_2014}, our algorithms use the more common approach that uses fixed step sizes. Similarly, while there exist \textit{early stopping} methods to identify the convergence of the reconstruction procedure \cite{yao_ca_2007}, we simply monitor the error metric $g(\OO,\PP)$ [Eq. \eqref{fitting_function}] for a fixed, large, number of iterations and choose the initial step sizes that give us the lowest values for the error metric.

To examine how the choice of these hyperparameters affects the reconstruction process at different experimental conditions, we followed the procedure in Section \ref{sec:results_farfield},
	but with the integrated probe intensity set to $10^3$ and $10^9$ photons respectively to generate two additional ptychographic datasets. To reconstruct the object and probe from each dataset, we used the Adam algorithm with a minibatch $b=100$, and performed a gridsearch on a logarithmic grid of initial probe update step sizes $[100, 10, 1, 0.1, 0.01]$ and initial object update step sizes of $[10, 1, 0.1, 0.01, 0.001]$. 
\begin{figure}[th]
	\centering
    \includegraphics[width=0.93\linewidth]{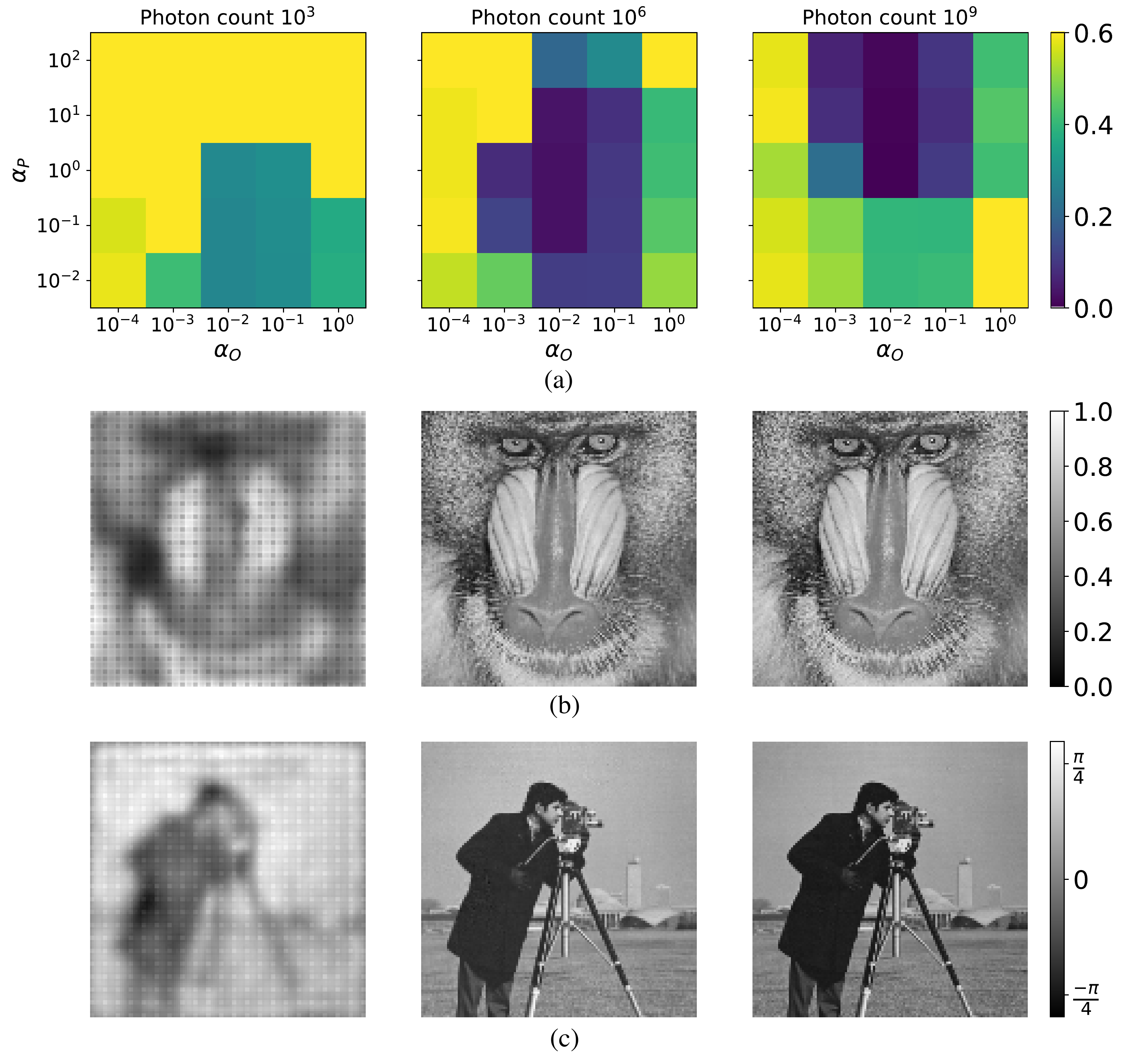}
	\caption{(a) Reconstruction errors (NRMSE) obtained after 1500 iterations of the Adam algorithm (with $b=100$)
		as a function of initial object and probe update step sizes, for incident probe integrated intensities of $10^3$ (left), $10^6$ (mid), and $10^9$ (right) respectively. For clarity, the NRMSEs plotted are capped at 0.6. (b) Final object magnitudes and (c) phases obtained for $\alpha_\sOO^A = 0.01$ and $\alpha_\sPP=1.0$. For the low photon count of $10^3$ (left), the reconstructed structures are deteriorated due to the raster grid artifact.}
	\label{fig:adam_hyperparameters}
\end{figure}

In \Cref{fig:adam_hyperparameters}, we show the object NRMSE obtained for each of these hyperparameter sets after 1500 iterations (a runtime of $\approx30 $s). We can see that $\alpha_\sOO^A=0.01$ remains an optimal object update step size for all three experiments. The probe update step size requires more careful tuning, with $\alpha_\sPP^A=1.0$ a good starting choice for the tuning procedure. For $(\alpha_\sOO^A,\alpha_\sPP^A)=(0.01, 1.0)$, after 1500 iterations, we obtained object and probe [$(\OO,\PP)$] reconstruction errors of $(0.28,0.51)$, $(0.03,0.02)$, and $(0.004,0.0004)$ respectively for integrated probe intensities of $10^3$, $10^6$, and $10^9$. In comparison, when we stop the ePIE algorithm ($b=1$) after 150 iterations ($\approx87$ s) without ensuring convergence, we get the  corresponding reconstructions errors of $(0.28, 0.50)$, $(0.06, 0.04)$, and $(0.05, 0.008)$. This demonstrates that the Adam algorithm, with the provided hyperparameter search strategy, is a fast and robust choice for ptychographic reconstruction within the AD framework.

\section{Applications to other ptychographic forward models}
\label{sec:generalizations}
In this section, we apply the reverse-mode AD framework developed in Section \ref{sec:ad_ptychography} 
for the far-field transmission ptychography experiment to the more complex experimental regimes
of near-field ptychography and multi-angle Bragg ptychography.

\subsection{Near-field ptychography}
\label{sec:results_near_field}
Just as in the far-field case, the near-field ptychography experiment uses a probe
beam to raster scan the object in a grid of spatially overlapping illumination spots, 
generating a set of $J$ exit waves after the probe-object interaction. The detector,
however, is placed at a distance $z$ from the object so that the Fresnel number satisfies the condition
\begin{align*}
	W^2  / \lambda z \gg 1,
\end{align*}
where $\lambda$ is the wavelength of the incident probe, 
and $W$ is the lateral extent of the illuminated area \cite{stockmar_scirep_2013}
at any given probe position. 
This is the high Fresnel number condition for the near-field
regime. In this regime, given a probe $\PP$, an object $\OO$, a lateral shift operator $\SSj$, and an exit wave $\psi_j = \PP\odot(\SSj\OO)$, 
the expected intensity at the detector plane is given by
\begin{align}
	h_j = \norm{D_z\{\psi_j\}}^2 + \nu_j
	\label{nearfield_intensity}
\end{align}
where $\nu_j$ is the expected background, and $D_z$ is the Fresnel free-space propagator for the distance $z$ defined by the expression
\begin{align}
	D_z\{\psi_j\} = \FF^{-1}\left[(\FF\psi_j )\odot 
								\exp{\frac{-i z \lambda}{4 \pi}\left(q_x^2 + q_y^2\right)} \right]
	\label{fresnel_propagator}
\end{align}
where $(q_x, q_y)$ the reciprocal space coordinates
\cite{goodman_fourier_2017}. The experimentally measured intensity patterns are again
denoted as $y_j$.
Detailed characterizations of this near-field ptychography
experiment can be found in \cite{stockmar_scirep_2013}, \cite{stockmar_pra_2015}, \cite{clare_oe_2015}, and \cite{robisch_njp_2015}.

We simulated a near-field ptychography experiment using an incident 
8.7 keV probe beam approximated using an array of size $512\times 512$ pixels at the object plane,
with the pixel pitch set to $0.6\,\mu$m.
The probe was initialized as a Gaussian with a FWHM of $19\,\mu$m (50 pixels),
then passed through a diffuser and modeled as a speckle pattern 
with an average flux of $10^4$ photons/pixel. 
The simulated object consists of the `Mandrill' and `Cameraman' images 
of size  $192\times 192$ pixels, modeled
as the object magnitude and phase respectively.
The raster grid was obtained by translating the object in the horizontal and vertical
in steps of $10\,\mu$m (44 pixels), generating a total of 25 diffraction patterns
at a detector placed $4.7$ cm from the object plane. 
We added Poisson noise to the 
diffraction patterns to obtain our simulated dataset.
\begin{figure}[th]
	\centering
	\begin{tabular}{cc}
		\includegraphics[width=0.31\linewidth]{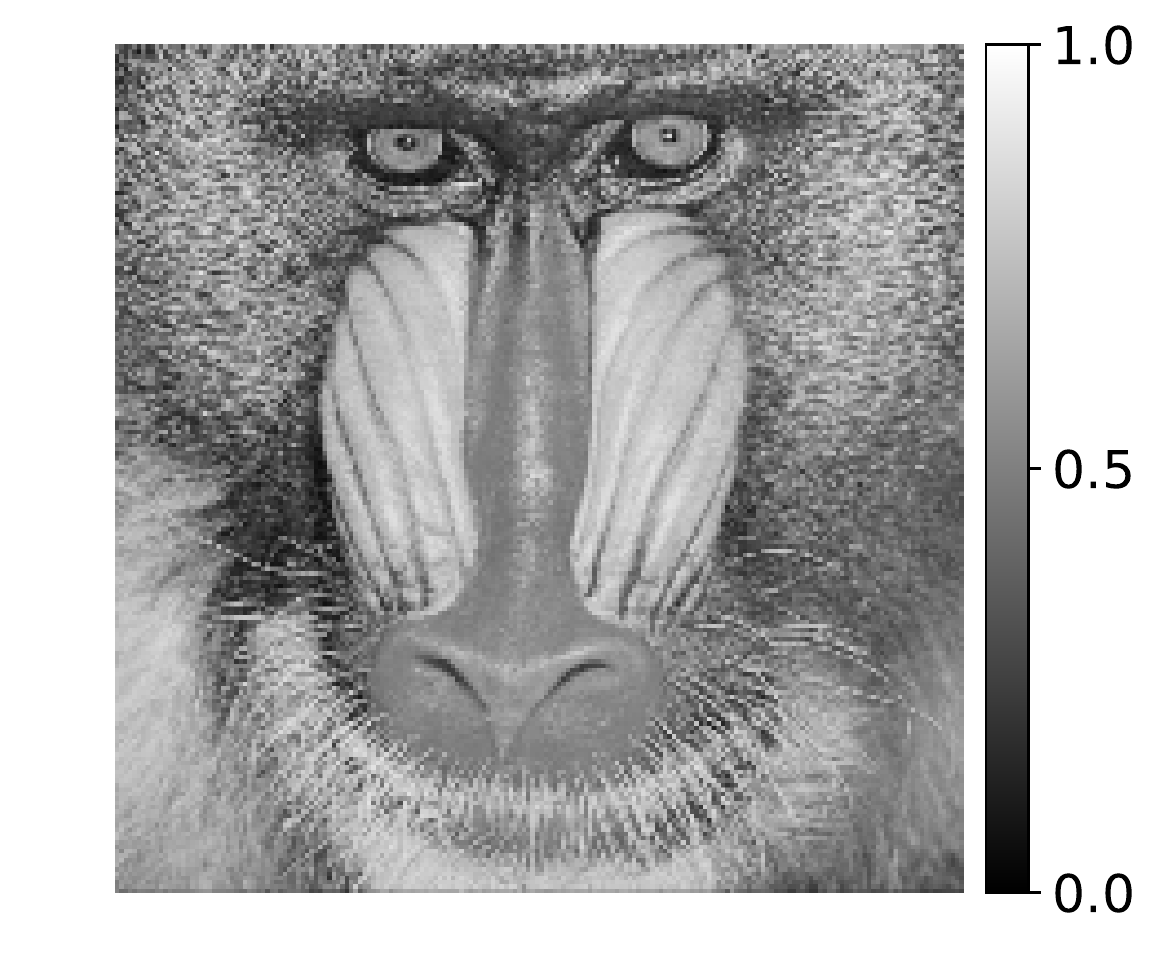} &
		\includegraphics[width=0.31\linewidth]{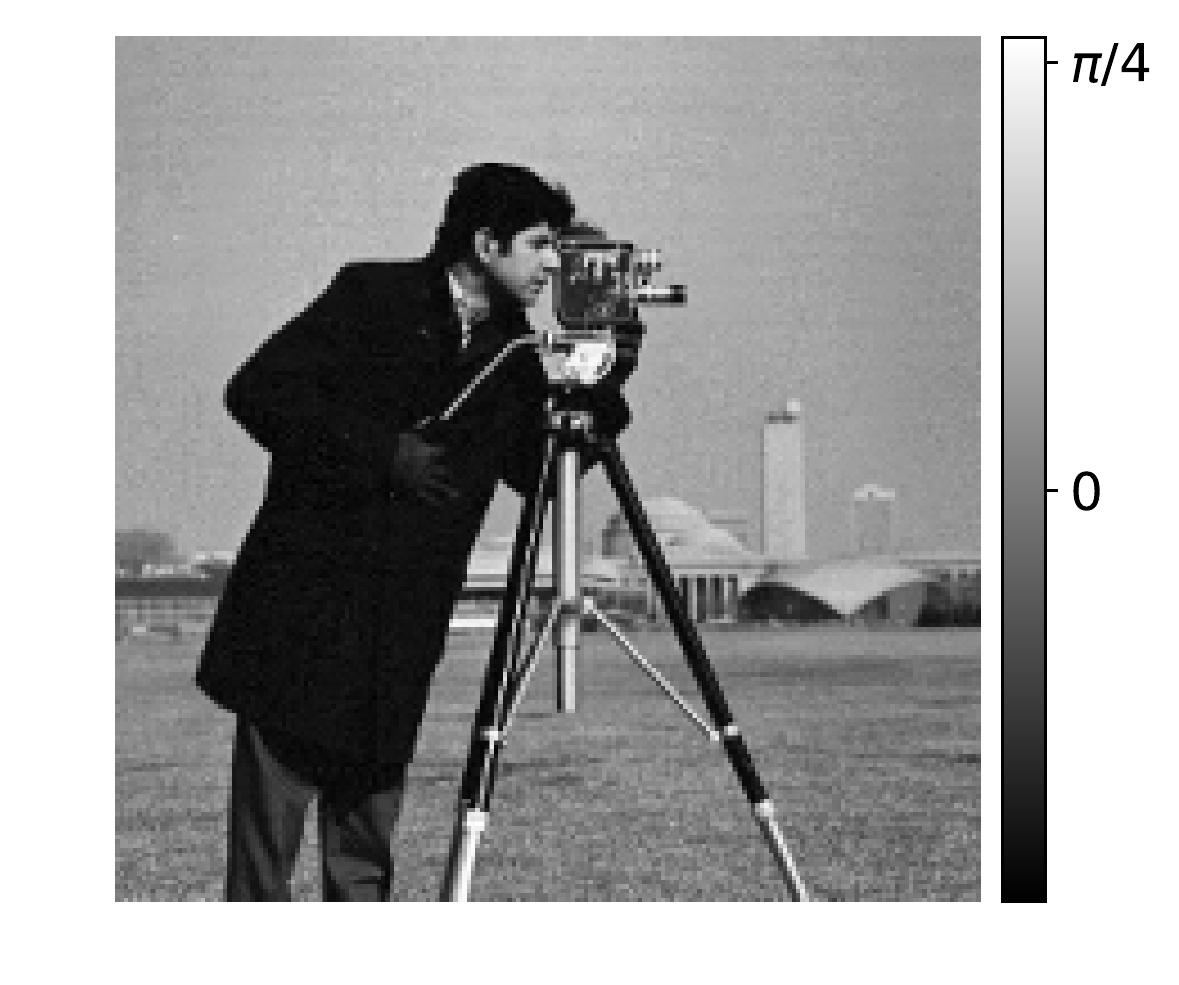} \\
		(a) & (b) \\
		\includegraphics[width=0.31\linewidth]{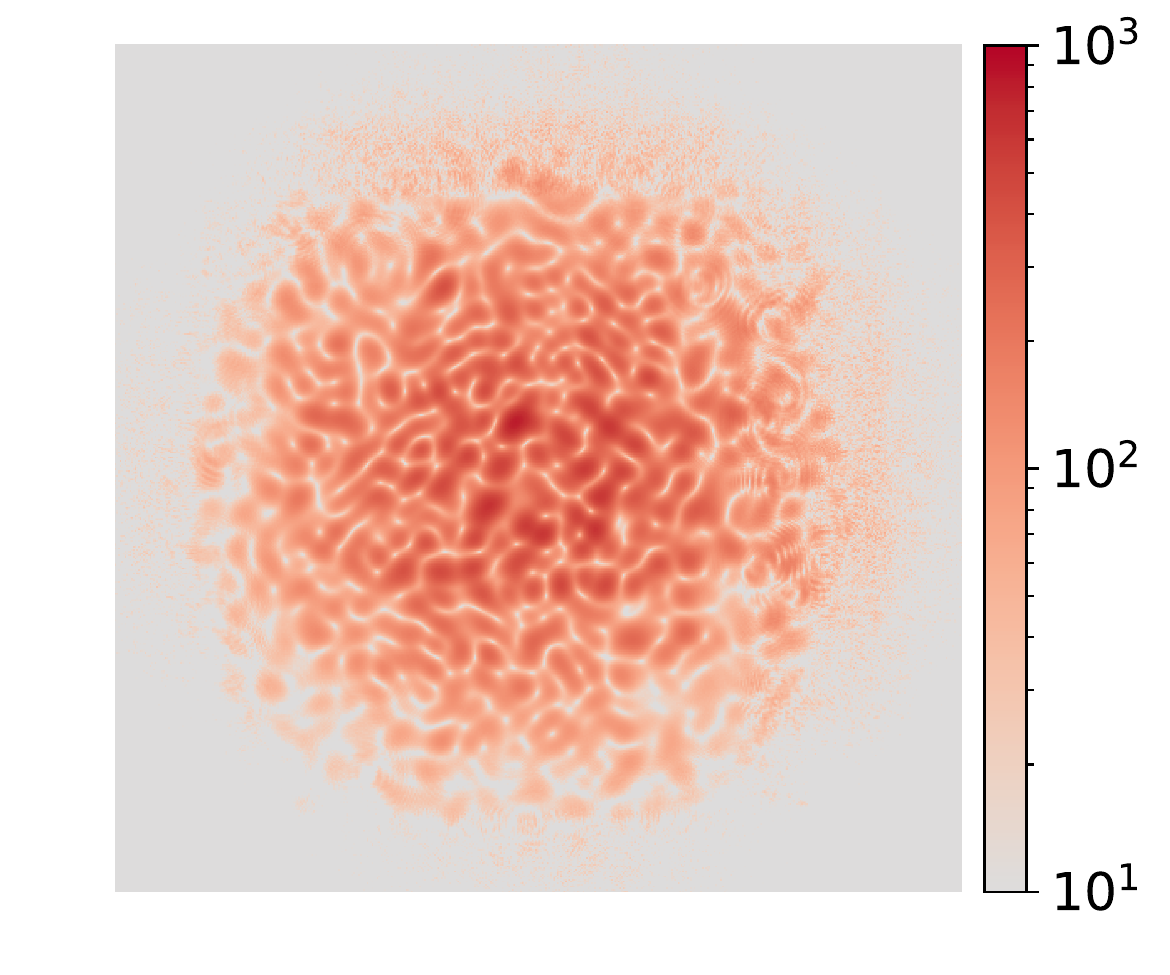} &
		\includegraphics[width=0.31\linewidth]{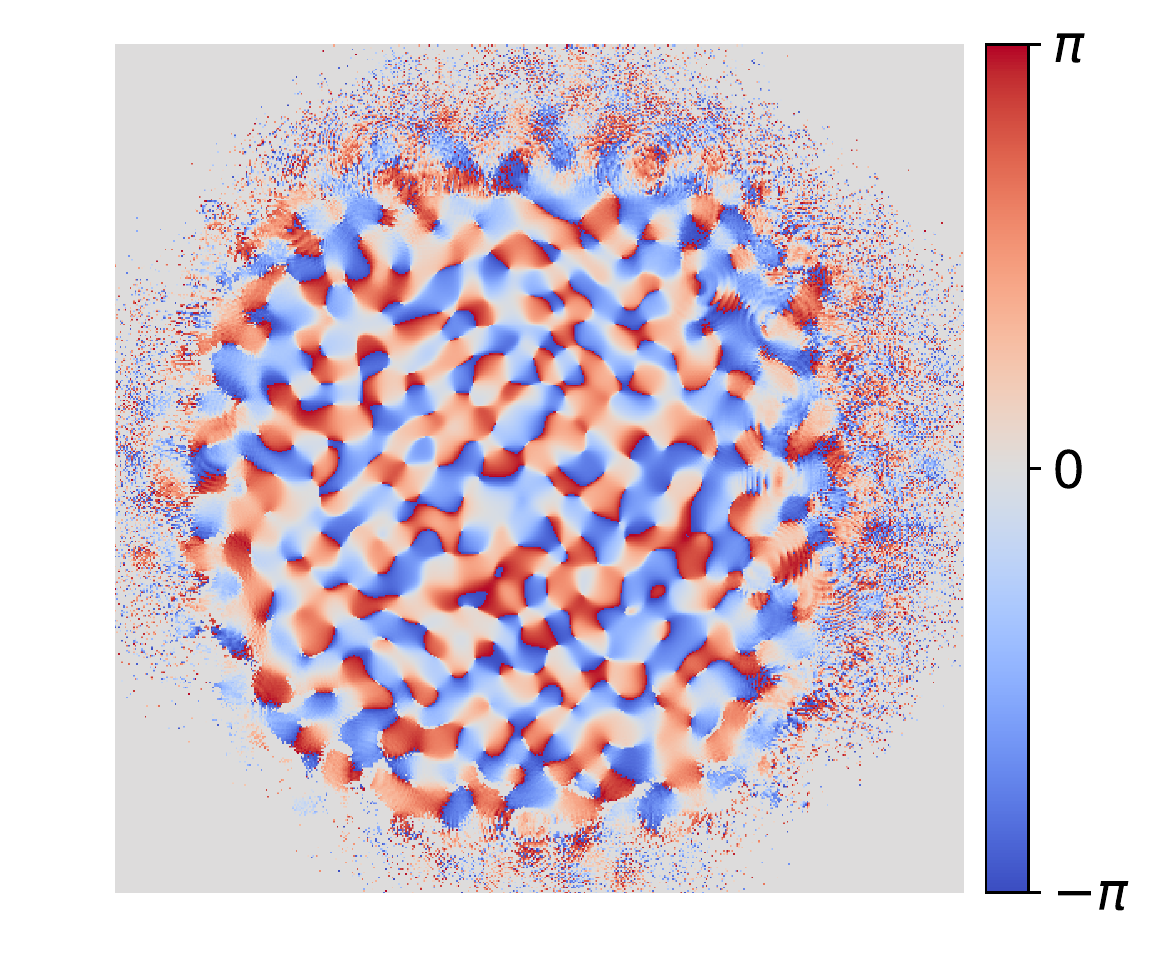} \\
		(c) & (d)
	\end{tabular}
	\caption{Near-field ptychographic reconstruction with reverse-mode AD. Shown here are the successfully reconstructed (a) object magnitude, (b) object phase, (c) probe magnitude, and (d) probe phase. The object and probe were reconstructed with overall NRMSEs of 0.01 and 0.12 respectively.}
	\label{fig:nearfield_results}
\end{figure}

For the near-field reconstruction, we obtained an initial probe estimate by backpropagating the average measured intensity, $y_{\text{avg}} = (\sumj y_j)/J$, to give
\begin{align}
\PP^0 = \FF^{-1}\left[(\FF\sqrt{y_{\text{avg}}})\odot\exp{\frac{iz\lambda}{4\pi}(q_x^2 + q_y^2)}\right].
\end{align}
The object was initialized as a $192\times192$ array of random complex numbers. 
To solve for the probe and object structures, we again used the Adam optimizer 
in the approach outlined
in Algorithm \ref{alg:epie_farfield}, but considering
5 probe positions per iteration (\textit{i.e.} with a minibatch size of 5).
The initial Adam update step sizes were set to $\alpha_\sOO^A=0.01$ and $\alpha_\sPP^A=10$
for the object and probe respectively. The minibatch size and initial step size were chosen as described in Section \ref{sec:hyperparameters}.

After 10,000 iterations of Adam gradient descent, we obtained the object and probe reconstructions 
shown in \Cref{fig:nearfield_results}. The object was reconstructed with an NRMSE of 0.01,
and the probe with an NRMSE of 0.12.
The discrepancies in the reconstructed probe were contained at the edges of the full-field probe, 
in regions which did not interact with the object and were thus unconstrained.
These reconstructions demonstrate that the straightforward generalization of the 
reverse-mode AD reconstruction framework from the far-field ptychography case ( \Cref{fig:farfield_graph}) 
to the near-field ptychography case
successfully solves the near-field ptychography phase retrieval problem.

\subsection{Multi-angle Bragg ptychography}
\label{sec:results_mabpp}
\begin{figure}[th]
	\centering
	\subfloat[]{
		\includegraphics[width=0.63\linewidth]{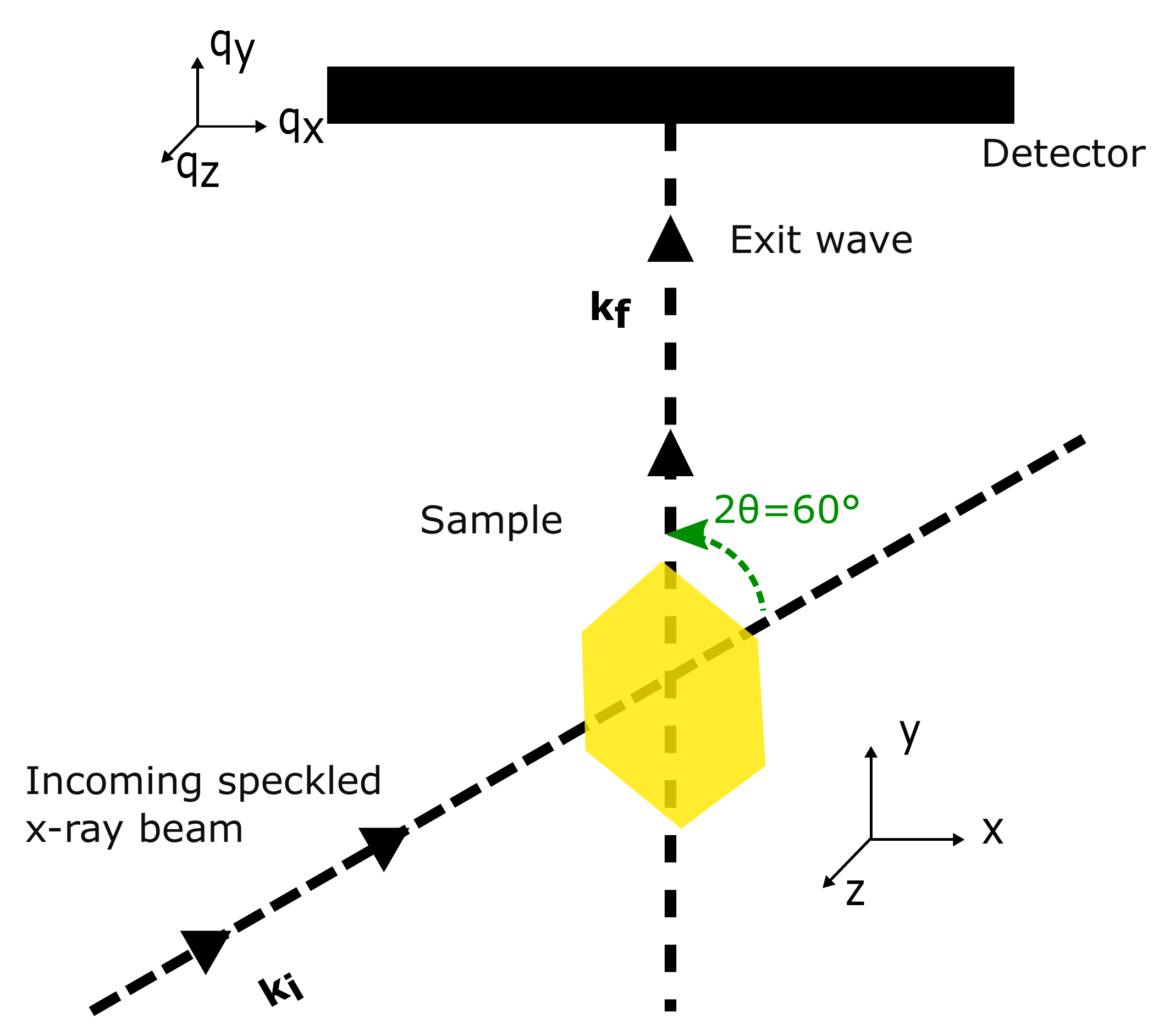}
	}
		\begin{minipage}[b]{0.3\linewidth}
			\subfloat[]{
				\includegraphics[width=0.9\linewidth]{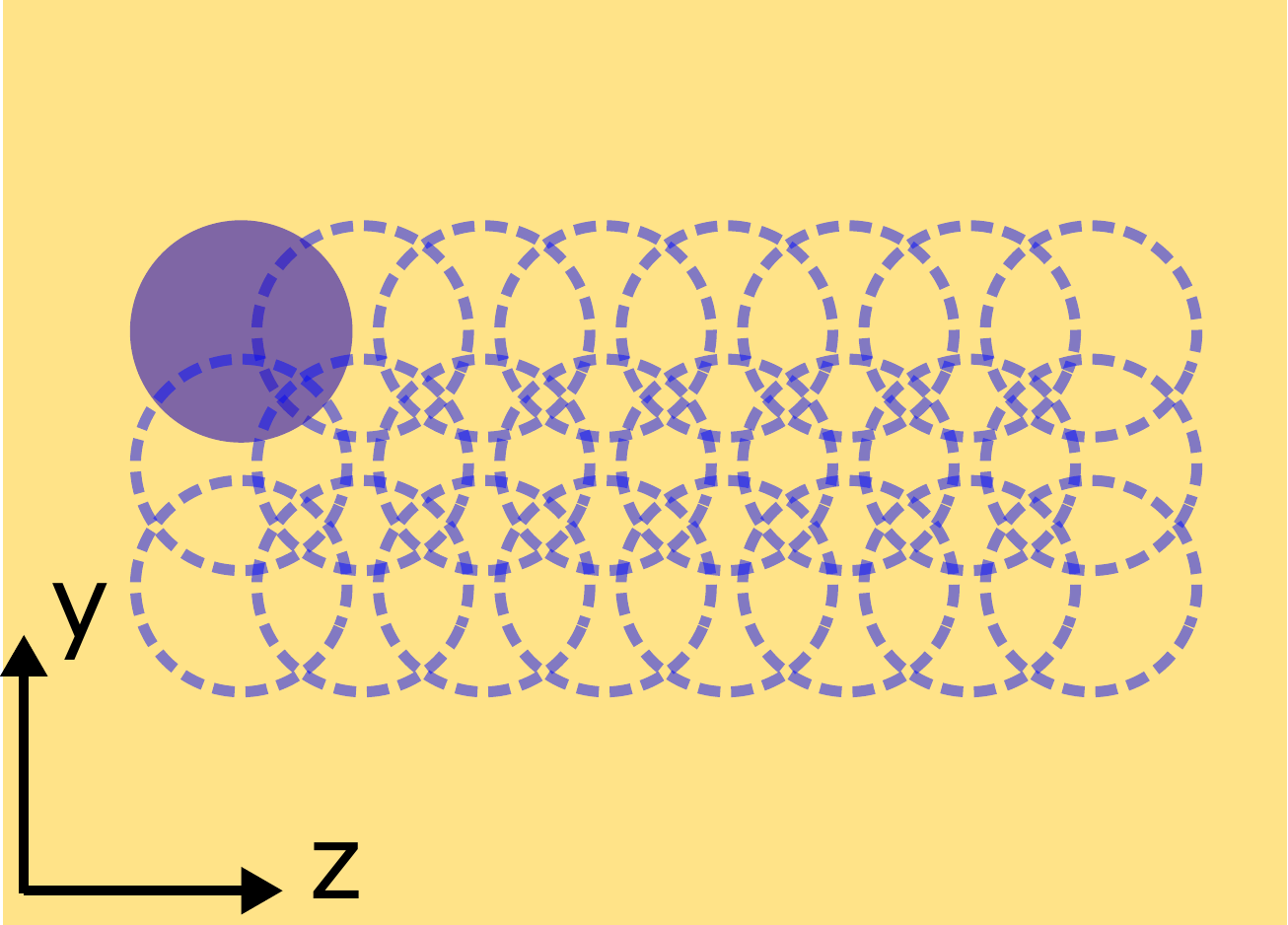}
			}\newline
			\subfloat[]{
				\includegraphics[width=0.9\linewidth]{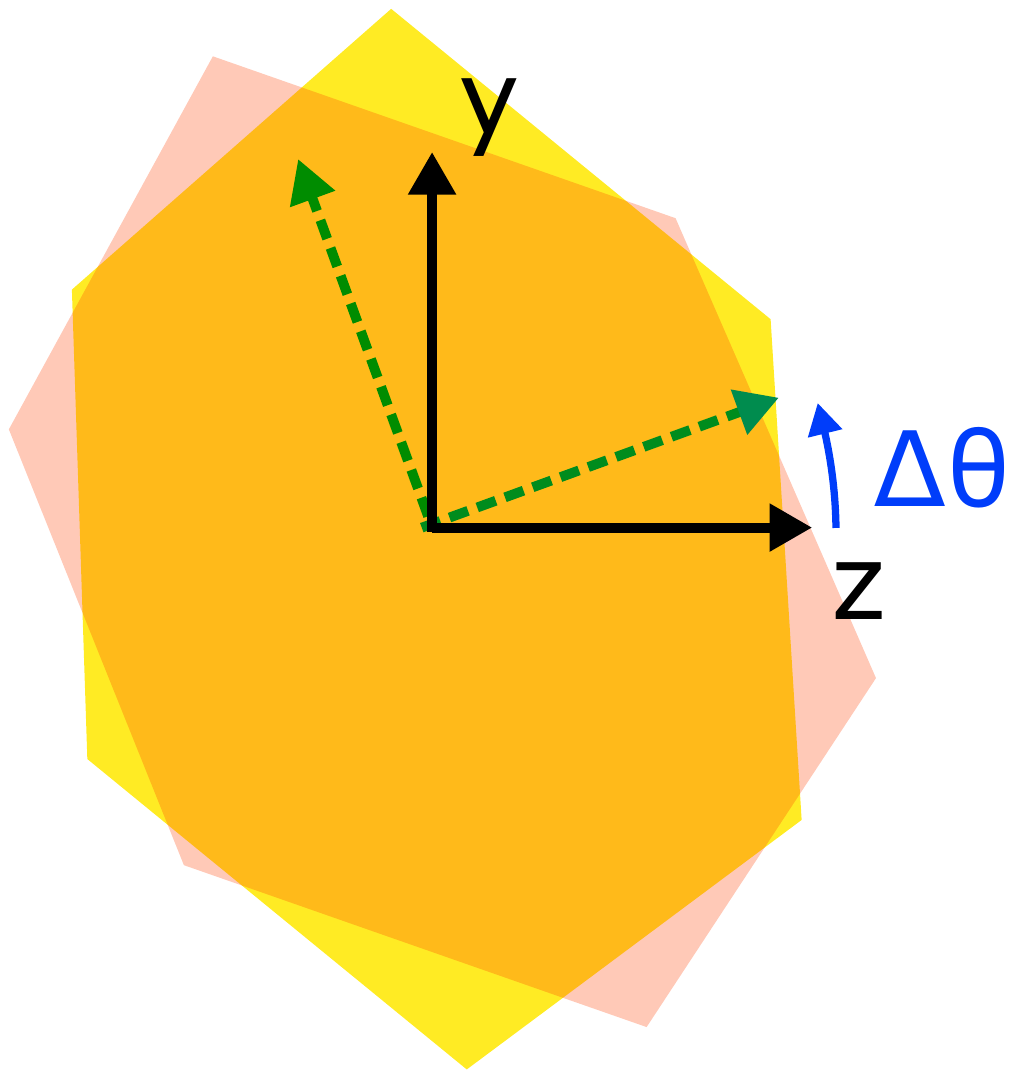}
			}
		\end{minipage}
	\caption{(a) Simulated experimental geometry for multi-angle Bragg ptychography. 
		The incident ($\mathbf{k}_i$) and exit ($\mathbf{k}_f$) beams define 
		a scanning angle. (b) At each scanning angle, the incident beam is shifted 
		along an overlapping raster grid in the $yz$ plane to generate a set of 
		2D coherent diffraction patterns. (c) For angular sampling around the 
		Bragg peak (say with $2\theta = 60^\circ$), the sample is rotated through 
		small angles $\Delta \theta$, and the incident and exit beams are modulated
		correspondingly.}
	\label{fig:mabpp_setup}
\end{figure}

Multi-angle Bragg projection ptychography (maBPP) \cite{hill_nl_2018} 
is a ptychographic experiment that allows for two degrees of freedom in the scan
parameters: 1) the choice of the planar scan positions for the usual two-dimensional
ptychographic scan, and 2) the choice of angular scan positions corresponding to 
small object rotations of a crystalline object oriented to satisfy a Bragg diffraction
condition. The maBPP experiment uses the far-field ptychography setup,
but with the detector placed to measure a crystalline Bragg peak, typically displaced from
the direct beam by tens of degrees. 
The detector records a set of two-dimensional (2D) coherent diffraction patterns at 
overlapping scan positions at one or more angles near the Bragg diffraction condition.
The diffraction patterns are then used to reconstruct a 
three-dimensional (3D) strain-sensitive image of an extended crystalline sample \cite{hruszkewycz_nm_2017}. An example experimental geometry with the 
exit beam at a $60^\circ$ angle
to the incident beam direction is shown in \cref{fig:mabpp_setup}.

To develop the maBPP forward model, we consider a 3D diffracting crystal volume
$\OO$ and an illumination volume $\PP$. The incident probe direction
$\ki$ and the exit beam direction $\kf$ satisfy the Bragg diffraction
condition when the scattering vector $\mathbf{q} = \kf - \ki$  coincides with
some reciprocal lattice vector $\mathbf{G}_{hkl}$ for the crystal. At the Bragg condition, if the probe volume interacts with the slice of the object indicated by the lateral shift operator $\SSj$, then the 2D exit wave $\psi_j$ is calculated as \cite{hruszkewycz_nm_2017}
\begin{align}
	\psi_j = \bm{\mathcal{R}}\PP\odot(\SSj\OO)
	\label{mabpp_bragg_exit_wave}
\end{align} 
where $\bm{\mathcal{R}}$ is the matrix operator performing the projection
along the exit beam direction. 

When the object is rotated by a small angle $\Delta \theta_j$, the scattering vector deviates
from the Bragg condition by $\mathbf{Q}_j = \mathbf{q}_j - \mathbf{G}_{hkl}$. 
The corresponding change in the object-probe interaction can be encoded 
in terms of a phase shift operator defined as 
$\mathfrak{Q}_j = \exp\left(i\mathbf{r}\cdot\mathbf{Q}_j\right)$. The 2D exit wave is then
given by \cite{hill_nl_2018}
\begin{align}
	\psi_j = \bm{\mathcal{R}}\mathfrak{Q}_j\odot\left(\PP\odot(\SSj\OO)\right),
\end{align} 
and the expected 
intensity at the detector plane is then 
\begin{align}%
	h_j = \norm{\FF\psi_j}^2 + \nu_j= \norm{\FF\bm{\mathcal{R}}\mathfrak{Q}_j\odot\left(\PP\odot(\SSj\OO)\right)}^2 + \nu_j,
	\label{3dpp_intensity}
\end{align}
where $\nu_j$ accounts for the background events. The experimentally measured intensities
are denoted by $y_j$.

We can again adapt the forward model of the far-field ptychography experiment to the
multi-angle Bragg experiment by generalizing the probe and object models to 3D, 
and applying the angle-dependent phase shifts $\mathfrak{Q}_j$ along with the 
projection operator $\bm{\mathcal{R}}$. This 3D generalization comes at the cost of a dramatic increase in the parameter space, 
and the reconstruction becomes correspondingly more difficult. With this in mind,
for ease of reconstruction 
we use a speckle pattern, which was found to help in solving the phase retrieval problem for 2D CDI \cite{fannjiang_ip_2012,edo_pra_2013}, as a highly diverse structured probe illumination.
We also impose the commonly applied \cite{hruszkewycz_nm_2017}
additional constraints on the reconstruction problem: we 
assume that the probe structure and the object thickness (along the $x-$direction) are known in advance.

For the multi-angle Bragg ptychography experiments, 
we simulated the object as a compact crystal represented by
set of grid points inside a faceted volume. 
We set the interior points to have magnitude 0.5 and a spatially varying phase structure emulating a strain field. We defined the orthonormal real space axes $(x,y,z)$ such that each object voxel is of size $66\times66\times66$ nm$^3$, and such that the polyhedral crystal is situated obliquely within (with $\approx32\%$ of the volume of) a cuboid of size $26\times56\times50$ voxels ($\approx25\,\mu\text{m}^3$).
The cuboid was itself placed at the center of a simulation box of total size $64\times162\times112$ voxels ($\approx334\,\mu\text{m}^3$). A $64\times64$ pixel detector with a pixel pitch of $55\,\mu$m was placed at a distance of 1.5 m from the object, normal to the real space $y$-axis. The probe was initialized as Gaussian beam of energy 8.7 keV with an FWHM of $396$ nm (6 pixels) contained entirely within a $64\times 64$ pixel array, then passed through a diffuser and modeled as a speckle pattern.
The probe was then interpolated to approximate the incident beam at 
a Bragg condition of $2\theta = 60^\circ$ (\cref{fig:mabpp_setup}) such that $\ki\perp z$ and $\kf\parallel y$. The beam profile was assumed to be constant along the propagation direction during its propagation through the simulation box. To obtain the ptychographic data set, we applied phase modulations corresponding to an angular shift between $\pm 0.14^\circ$ around the Bragg condition using an angular step size of $\Delta\theta=0.02^\circ$, for a total of 15 angles. At each such angle, we translated the probe along the $y$ and $z$ directions with a step size of 132 nm (2 pixels) in a raster grid of $41\times24$ scan positions, \textit{i.e.} with an overlap of $\approx86\%$ per step in the $y$ direction and $\approx 83\%$ per step in the $z$ direction. This generated a total of 14760 diffraction patterns, to each of which we added Poisson noise, which gave us a data set with an overall intensity maximum of 13560 photons/pixel.
\begin{figure}[th]
	\centering
	\subfloat[]{
		\includegraphics[width=0.21\linewidth]{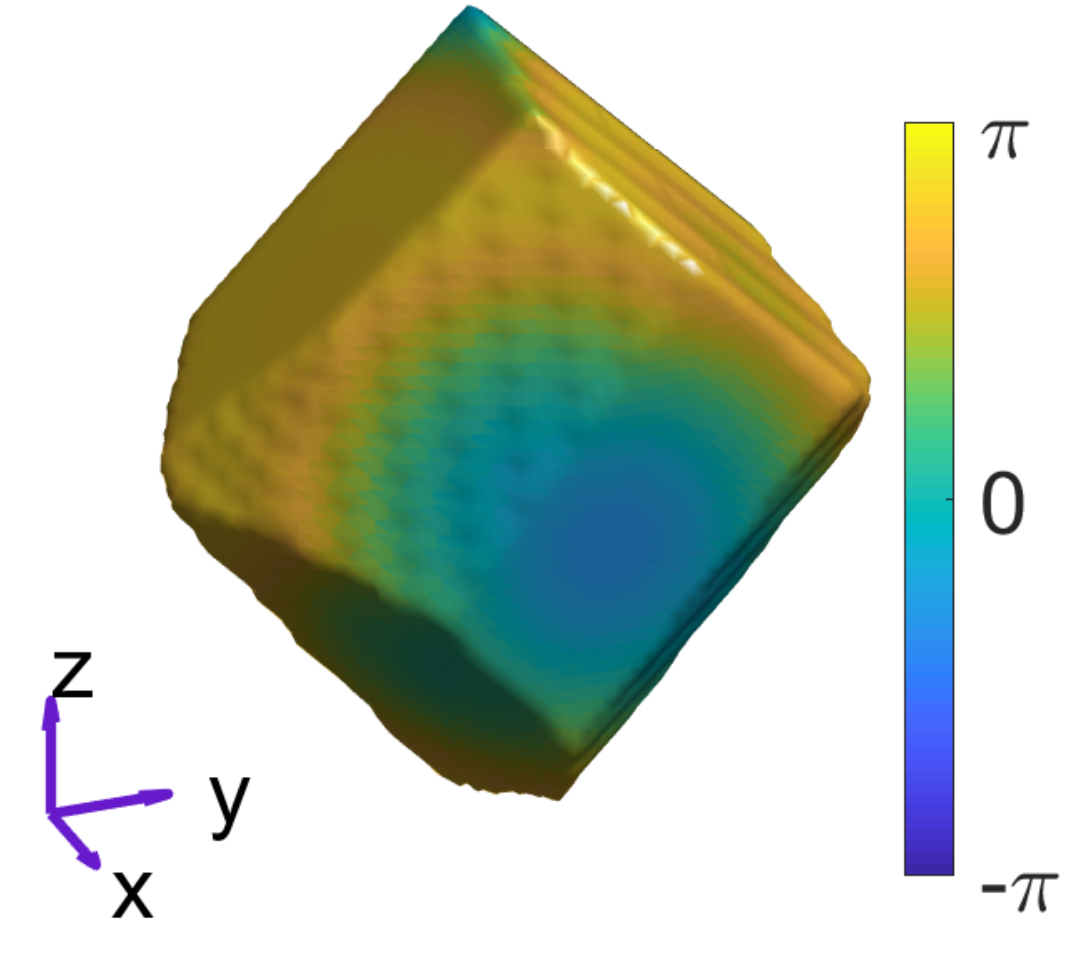}
	}
	\subfloat[]{
		\includegraphics[width=0.21\linewidth]{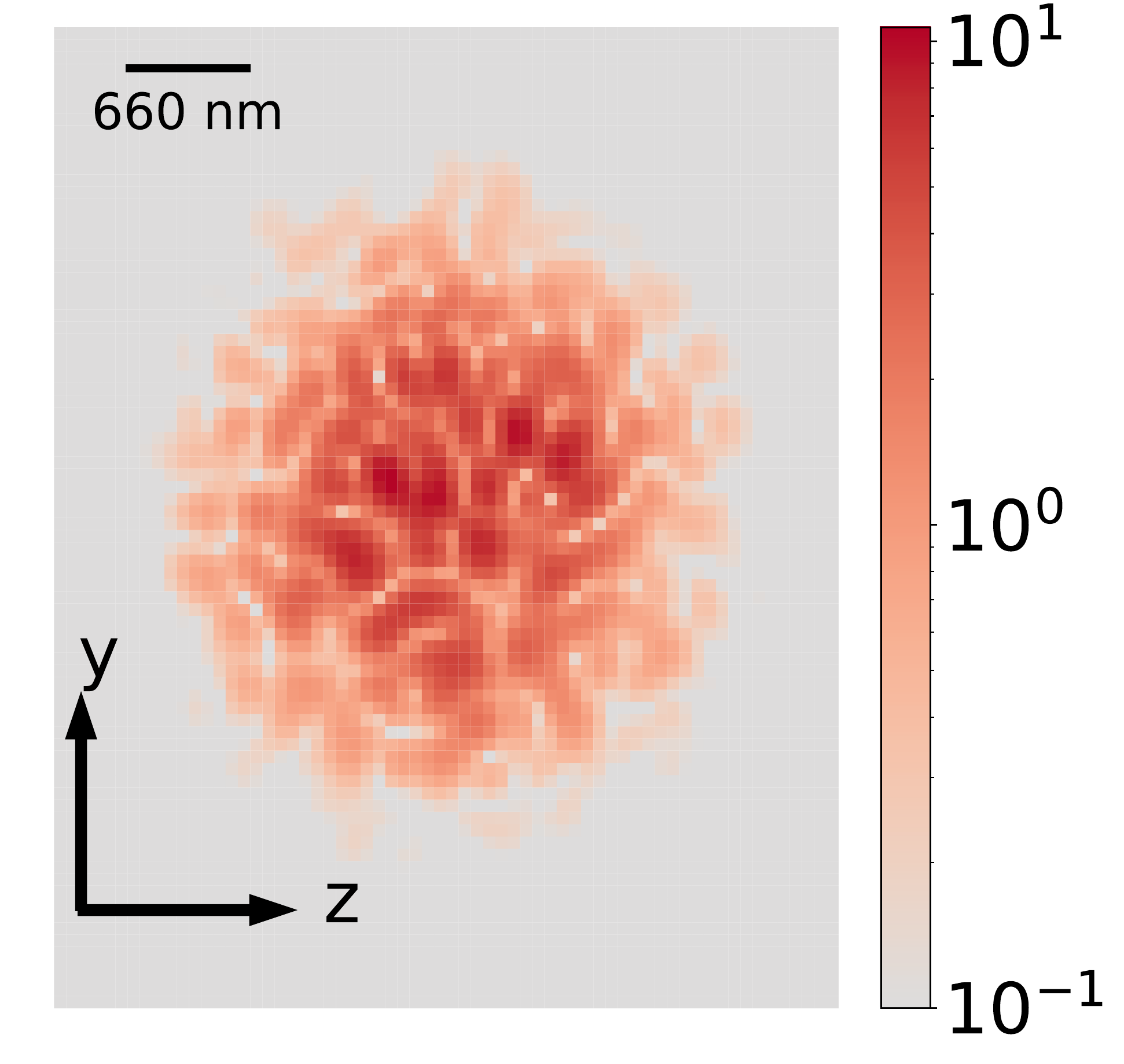}
	}
	\subfloat[]{
		\includegraphics[width=0.51\linewidth]{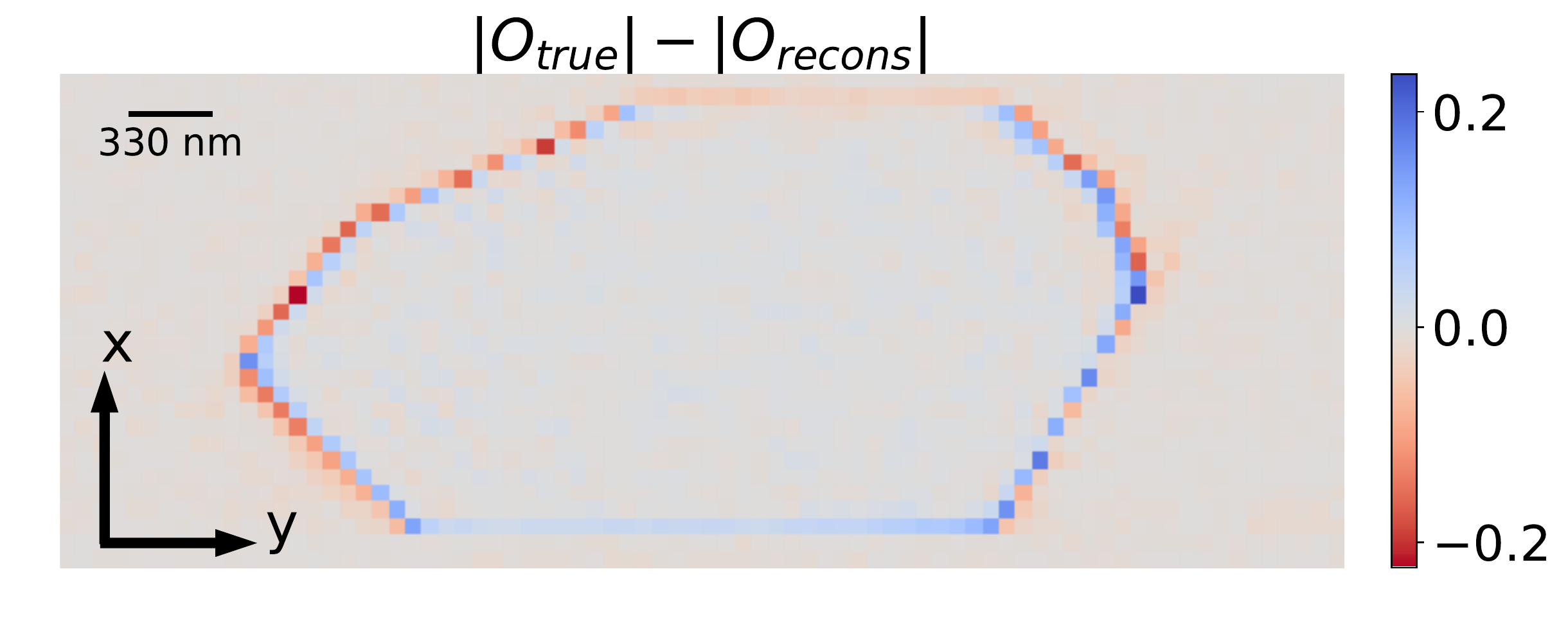}
	}\newline
	\subfloat[]{
		\includegraphics[width=0.21\linewidth]{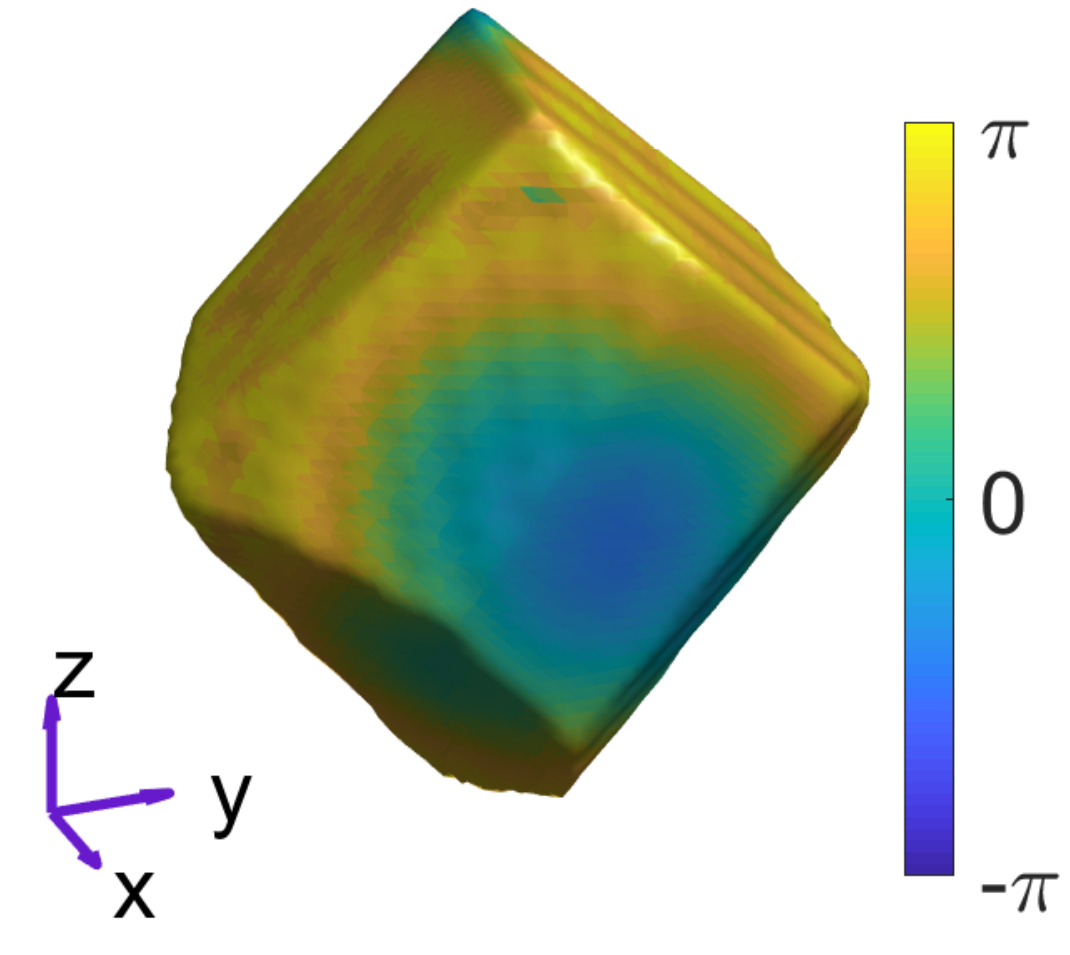}
	}
	\subfloat[]{
		\includegraphics[width=0.21\linewidth]{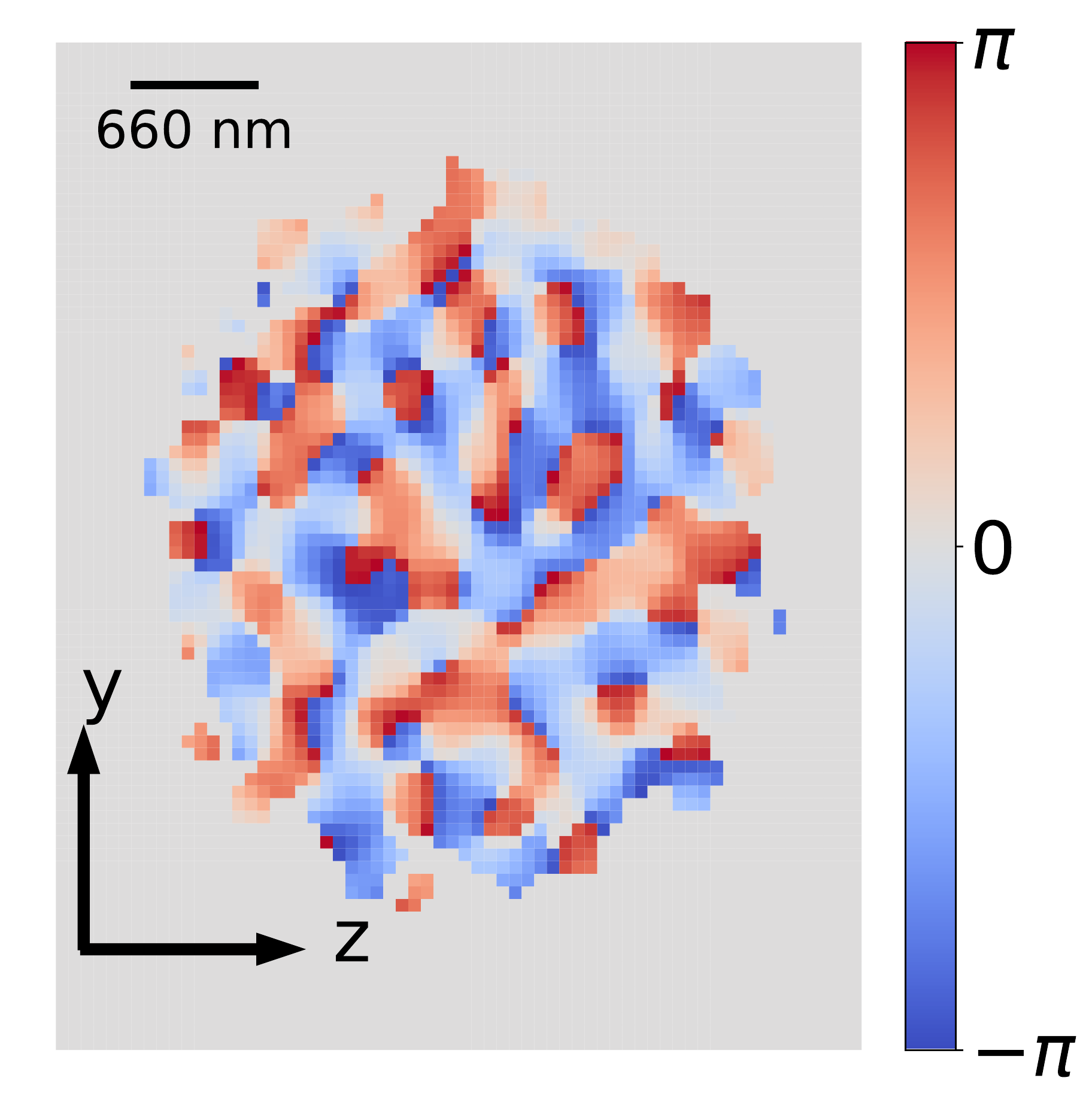}	
	}
	\subfloat[]{
		\includegraphics[width=0.51\linewidth]{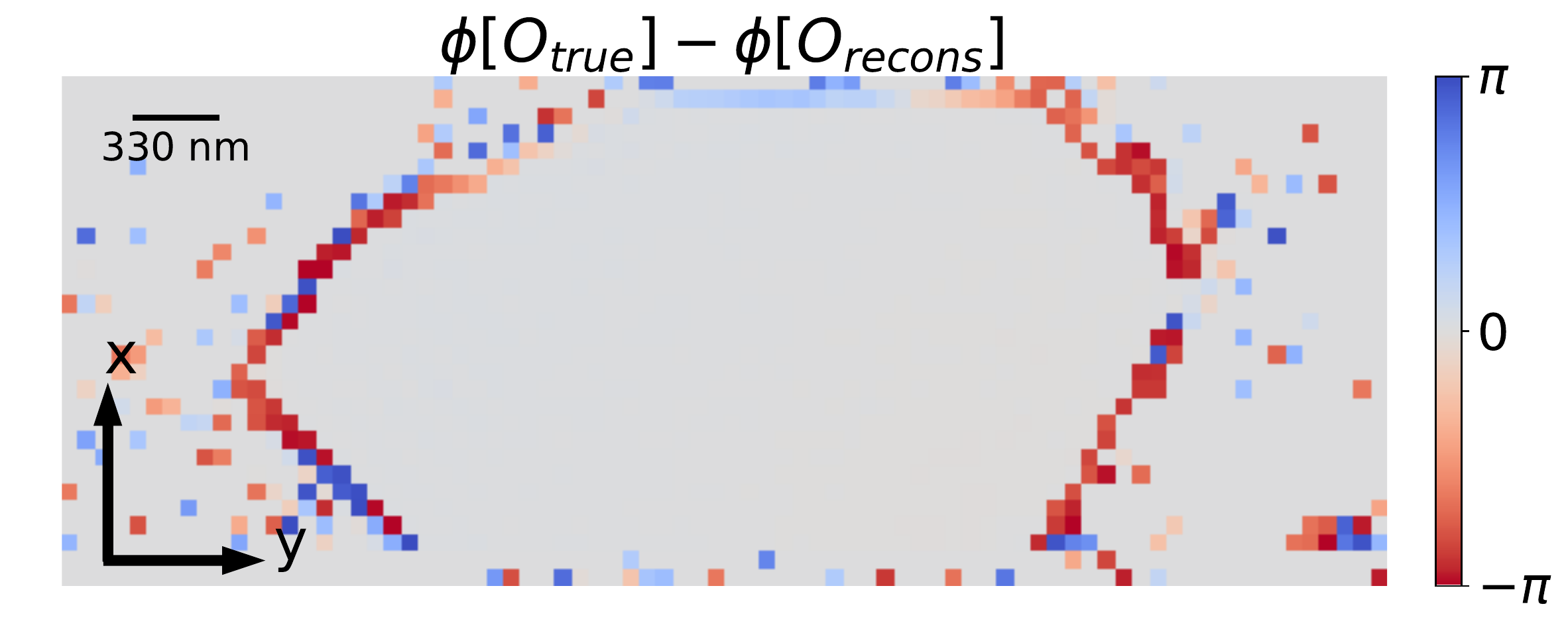}
	}
	\caption{Multi-angle Bragg ptychographic reconstruction with reverse-mode AD. (a) True and (d) reconstructed object structure with surface phase variation.
		$YZ$ cross-section of the (b) probe magnitude and (e) phase at $x=32$. (c) $XY$ cross-section of the error in the reconstructions of the object (c) magnitude and (f) phase at $z=41$. For the sake of clarity, the error in the phase is set to 0 wherever $\norm{\OO_{recons}} < 0.01$. The normalized error (NRMSE) for the 3D reconstruction is 0.09.}
		\label{fig:mabpp_results}
\end{figure}

During the reconstruction, we solved for a volume of size $30\times78\times82$ voxels, initialized as a random complex array, and placed centrally in the simulation box. The object support along the $x-$direction is set to be slightly ($\approx15\%$) larger than the actual object thickness. This allows for the expected spreading of the object induced by resolution effects due to the low photon count \cite{chamard_sr_2015}. 
To solve for the object structure, we again adapted the approach outlined in 
Algorithm \ref{alg:epie_farfield} to use the Adam optimizer with the initial update step size $0.01$, 
and with minibatches each containing 400 diffraction patterns 
selected in a stochastic fashion irrespective of either probe position or object rotation.
The minibatch size and initial update step size were chosen as described in Section \ref{sec:hyperparameters}. After 70 iterations of Adam gradient descent,
we obtained a reconstruction of the crystal (and the loose support) shown in \Cref{fig:mabpp_results}
The overall normalized reconstruction error, 
calculated using a 3D adaptation of the sub-pixel registration algorithm presented in \cite{guizar_ol_2008}, 
was found to be 0.09, which is much larger than the NRMSE observed for the 2D objects in Sections \ref{sec:results_farfield}
and \ref{sec:results_near_field}. As we can see in \Cref{fig:mabpp_results}, 
the error in the reconstruction is primarily due to discrepancies at the object edges---this is a physical effect that
can be attributed to the presence of shot noise that corrupts the diffraction data \cite{chamard_sr_2015}. 
This results demonstrates that the reconstruction accurately captures the physics of the experiment.

\section{Conclusion}
Our results in this paper demonstrate that the automatic differentiation technique
can be used to construct a general gradient-based inversion framework for ptychographic 
phase retrieval. Specifically, we have shown that we can minimize a ptychographic error metric 
by first using reverse-mode AD to numerically calculate the gradients of the error metric, 
then using these within a state-of-the-art adaptive gradient descent
algorithm.  We can thereby solve the ptychographic problem without ever performing an analytical derivation for any closed-form gradient expression. This inversion framework is robust to changes
in the forward model and can be used identically for phase retrieval in such varying
experimental configurations as far-field transmission ptychography, near-field ptychography,
and multi-angle Bragg ptychography.

The inversion framework showcased in this work does not rely on any particular choice of 
the error metric or the experimental configuration. 
We can change the error metric either to accommodate alternate noise models,
or to incorporate \textit{a priori} knowledge about the experiment through the choice
of a regularizer. We can tailor the experimental setup and introduce new degrees of freedom
when necessary, then simply change the forward model correspondingly. Additionally, by using the state-of-the-art TensorFlow library for the gradient calculations, we gain in-built scalability and parallelization (multi-CPU/GPU architectures), thus allowing for convenient phase retrieval for experimental models large and small. In the future, we aim to leverage these flexibilities to incorporate a variety of physical phenomena--such as probe fluctuations, positional inaccuracies, and limitations in the detection process--into the phase retrieval framework. This would enable high-resolution phase retrieval from near-field/far-field 3D (or 2D) Bragg and non-Bragg experiments, all within a single consistent platform.

The flexible AD-based inversion framework is expected to provide a unified approach to phase retrieval for general (ptychographic and non-ptychographic) CDI experiments with x-rays, electrons, or optical waves, even as we move towards increasingly complex imaging regimes. This would potentially give researchers the ability to explore variations of basic CDI methodologies in a convenient and straightforward manner, and could thereby prove a powerful addition to the phase retreiver's toolbox. 

\appendix
\section{Parameter update with the Adam optimizer}
\label{sec:adam_alg}

To examine how the Adam optimizer performs the parameter updates, we consider the 
far-field transmission ptychography setting from Section \ref{sec:forward_model_farfield}, 
with the error metric $g(\OO, \PP)$ which is to be minimized with respect to the object parameter $\OO$. The optimization is performed in the minibatch setting described in 
Algorithm \ref{alg:epie_farfield}, wherein the partial derivative at the $k^{th}$ step,
$\delO g^k$, is calculated using Eq. \eqref{minibatch_grads}. 

The Adam optimization step consists of a parameter initialization step and a variable 
update step. The initialization step, which precedes the gradient descent iterations,
is outlined in Algorithm \ref{alg:adam_init}.  The parameter update step in Algorithm 
\ref{alg:adam_update} replaces Steps 4 and 5 in the iterative descent setting
outlined in Algorithm \ref{alg:epie_farfield}. Typical out-of-the-box implementations
of the Adam optimizer only support gradient descent for real-valued variables, 
in which case the real and imaginary components of the object 
$\OO$ are separately updated. 
To formalize this update step, we can separate the real and imaginary components of $\delO g^k$ as 
\begin{align}
\boldsymbol{\partial}_{_{\Rel{\OO}}} g^k = \frac{1}{2}\pdv{g^k}{\Rel{\OO}}\quad
\text{and}\quad
\boldsymbol{\partial}_{_{\Img{\OO}}} g^k = \frac{1}{2}\pdv{g^k}{\Img{\OO}}
\label{derivative_wrt_real_img_O}
\end{align}
such that $\delO g^k = \boldsymbol{\partial}_{_{\Rel{\OO}}} g^k + i \boldsymbol{\partial}_{_{\Img{\OO}}} g^k$.
We can now see that the first and second moment parameters created during the initialization
step for $\Rel{\OO}$ (and similarly for $\Img{\OO}$) are subsequently updated alongside the object
variable throughout the optimization process. These updates depend on both
the current gradient value and an exponentially weighted accumulation of all past gradient
values, thereby adapting the magnitude of direction of the parameter update 
at every step. For an appropriately chosen initial step size $\alpha_\sOO^A$, updating
the optimization parameter in this fashion leads to fast and stable convergence \cite{kingma_corr_2014}.
\begin{subalgorithms}
	\begin{figure*}[!htb]
		\hspace*{\fill}
		\begin{minipage}[t]{0.45\linewidth}
			\begin{algorithm}[H]
				\caption{Adam object initialization}
				\label{alg:adam_init}
				\begin{algorithmic}[1]
					\Require Initial Adam step size $\alpha^A_\sOO$.
					\State Initialize the moment vectors
					\begin{align*}
					&\left.\begin{aligned}
					m_{_\RO}^0\longleftarrow 0\\
					m_{_\IO}^0\longleftarrow 0
					\end{aligned}\right\}
					\left.\begin{aligned}
					\text{First moments,}\\
					m_{_\RO}, m_{_\IO}\in\Reals^N
					\end{aligned}\right.\\
					&\left.\begin{aligned}
					v_{_\RO}^0\longleftarrow 0\\
					v_{_\IO}^0\longleftarrow 0
					\end{aligned}\right\}
					\left.\begin{aligned}
					\text{Second moments,}\\
					v_{_\RO}, v_{_\IO}\in\Reals^N
					\end{aligned}\right.\\
					\end{align*}
					\State Initialize the exponential decay rates with their recommended default values \cite{kingma_corr_2014}:
					\begin{align*}
					&\beta_1 = 0.9 &(\text{Decay rate for }m_{_\RO},m_{_\IO})\\
					&\beta_2 = 0.999 &(\text{Decay rate for }v_{_\RO},v_{_\IO})\\
					&\epsilon = 10^{-8} &(\text{Constant used to avoid}\\
					&&\text{division by zero})
					\end{align*}
				\end{algorithmic}
			\end{algorithm}
		\end{minipage}
		\hfill
		\begin{minipage}[t]{0.50\linewidth}
			\begin{algorithm}[H]
				\caption{Adam object update (at step $k$)}
				\label{alg:adam_update}
				\begin{algorithmic}[1]
					\Require Object estimate $\OO^{k-1}$ and the partial derivative $\delO g^k$.
					\State Update the first and second moments:
					\begin{align*}
					m_{_\RO}^k &= \frac{\beta_1 m_{_\RO}^{k-1} + 
						(1 - \beta_1) \delRO g^k}{1-(\beta_1)^k}\\
					v_{_\RO}^k &= \frac{\beta_2 v_{_\RO}^{k-1} + (1 - \beta_2) \norm{\delRO g^k}^2}{1-(\beta_2)^k}
					\end{align*}
					(and similarly  for $m_{_\IO}^k$ and $v_{_\IO}^k$).
					\State Update the object estimate:
					\begin{align*}
					\RO^k &= \RO^{k-1} - \alpha_\sOO^A\frac{m_{_\RO}^k}{\sqrt{v_{_\RO}^k} + \epsilon}\\
					\IO^k &= \IO^{k-1} - \alpha_\sOO^A\frac{m_{_\IO}^k}{\sqrt{v_{_\IO}^k} + \epsilon}
					\end{align*}
				\end{algorithmic}
			\end{algorithm}
		\end{minipage}
		\hspace*{\fill}
	\end{figure*}
	\label{alg:adam}
\end{subalgorithms}
\FloatBarrier

\section*{Funding}
We thank the following for funding of this work: Laboratory Directed Research and Development (LDRD) and the Advanced Photon Source, Argonne National Laboratory, provided by the Director, Office of Science, of the U.S. Department of Energy under Contract No. DE-AC02-06CH11357;
National Institutes of Health under R01 MH115265; and the
Department of Energy, Office of Science, Basic Energy Sciences, Materials Science and Engineering Division.

\clearpage

\bibliography{xrmbook}
\bibliographystyle{jabbrv_ieeetr}

\end{document}